\documentclass[10pt]{article}

\usepackage{microtype}
\usepackage{graphicx}
\usepackage{subcaption}
\usepackage{booktabs} 

\usepackage{hyperref}
\usepackage{xurl} 

\providecommand{\doi}[1]{\href{https://doi.org/\detokenize{#1}}{\nolinkurl{doi:#1}}}

\usepackage[preprint]{icml2026}

\usepackage{colortbl}
\pdfoutput=1
\usepackage{amsmath, amssymb, amsthm}
\usepackage{mathtools}
\usepackage{enumitem}
\usepackage{balance}
\usepackage{xcolor}
\usepackage{listings} 
\usepackage{tabularx}
\usepackage{makecell}
\usepackage{framed}   
\usepackage{caption}  
\usepackage{needspace} 
\usepackage[section]{placeins} 

\newcommand{\lightningsym}{\textbf{!}}

\lstdefinelanguage{json}{
    basicstyle=\ttfamily\small,
    breaklines=true,
    frame=single,
    keywordstyle=\bfseries\color{blue},
    morestring=[b]",
    morecomment=[l]{//},
    literate=
     *{true}{{\color{green}true}}{4}
      {false}{{\color{red}false}}{5}
      {null}{{\color{gray}null}}{4}
}

\newcommand{\prompttight}{%
  \setlength{\parindent}{0pt}%
}

\begin{document}

\twocolumn[
  \icmltitle{RAGXplain: From Explainable Evaluation to Actionable Guidance of RAG Pipelines}
  \icmlsetsymbol{equal}{*}

    \begin{icmlauthorlist}
    \icmlauthor{Dvir Cohen}{equal,wix}
    \icmlauthor{Tamir Houri}{equal,wix}
    \icmlauthor{Lin Burg}{wix}
    \icmlauthor{Gilad Barkan}{wix}
    \end{icmlauthorlist}
  
    \icmlaffiliation{wix}{Wix.com AI Research, Tel Aviv, Israel}
  \icmlcorrespondingauthor{Dvir Cohen}{dvirco@wix.com}

  \icmlkeywords{Machine Learning, ICML}

  \vskip 0.3in
]

\printAffiliationsAndNotice{\icmlEqualContribution}  


\begin{abstract}
Retrieval-Augmented Generation (RAG) systems couple large language models with external knowledge, yet most evaluation methods report aggregate scores that reveal \emph{whether} a pipeline underperforms but not \emph{where} or \emph{why}.
We introduce \textbf{RAGXplain}, an evaluation framework that translates performance metrics into actionable guidance.
RAGXplain structures evaluation around a ``Metric Diamond'' connecting user input, retrieved context, generated answer, and (when available) ground truth via six diagnostic dimensions.
It uses LLM reasoning to produce natural-language failure-mode explanations and prioritized interventions.
Across five QA benchmarks, applying RAGXplain’s recommendations in a single human-guided pass consistently improves RAG pipeline performance across multiple metrics.
We release RAGXplain as open source to support reproducibility and community adoption.
\end{abstract}

\section{Introduction}
\label{sec:introduction}

RAG systems integrate LLMs with external knowledge retrieval to produce more factual, relevant, and contextually grounded outputs. This paradigm's utility spans a wide spectrum, from enhancing the timeliness (i.e., freshness) of web search \cite{nakano2022webgptbrowserassistedquestionansweringhuman,Wang_2023} to supporting high-stakes applications in finance, law, and biomedicine that demand high fidelity and transparency \cite{li2024experimentinglegalaisolutions, xiong2024benchmarkingretrievalaugmentedgenerationmedicine, kim2025optimizingretrievalstrategiesfinancial}. 

However, rigorous evaluation remains difficult and often non-actionable: end-to-end scores (e.g., relevance or faithfulness) rarely reveal \textit{where} the pipeline failed (retrieval vs.\ generation) or \textit{why} (missing evidence, off-topic context, hallucinated details, formatting gaps, or other failure modes).
This diagnostic ambiguity is especially limiting for practitioners without specialized machine learning expertise, who need interpretable, failure-mode-specific feedback rather than aggregate scalar scores to decide what to change.

\begin{figure}[!t]
    \centering
    \includegraphics[width=0.90\linewidth]{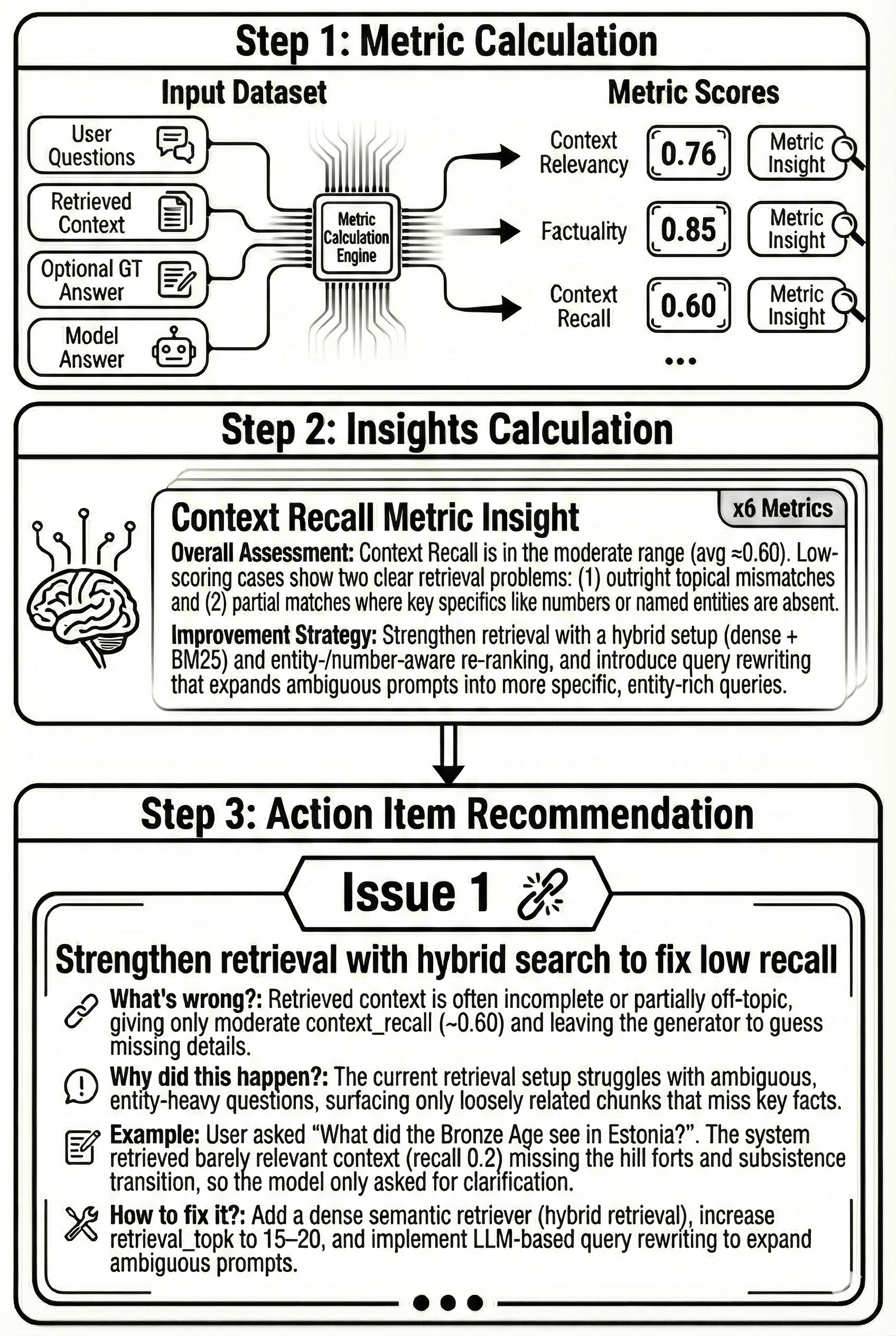}
    \caption{RAGXplain converts metric outputs into human-readable diagnoses and concrete action items.}    \label{fig:pipeline}
\end{figure}

Many RAG evaluation frameworks report quantitative scores for dimensions such as relevance, faithfulness, and correctness \cite{ragas,ares,vrag_eval}, with some providing more fine-grained or stage-aware analyses \cite{ragbench,cofe_rag}. 
While valuable for benchmarking, these aggregate scores often provide limited actionable guidance. Recent work also shows that even strong automated evaluators can exhibit reliability gaps and blind spots that a single aggregate score cannot diagnose \cite{muller2025grousebenchmarkevaluateevaluators, liu2025judgejudgeimprovingevaluation}.
This challenge is compounded by RAG's high-dimensional, multi-stage design.
Beyond the base LLM, practitioners must configure document processing (e.g., chunking and overlap), embedding models and indexing, retrieval and reranking, and generation prompts/decoding where each component requires task-specific hyperparameters.
Because these components interact, retrieval and generation errors can mask one another, making it difficult even for technical teams to efficiently pinpoint whether a failure originates in retrieval (e.g., missing evidence) or generation (e.g., hallucination).

RAGXplain connects quantitative assessment to concrete optimization steps. It computes a customizable suite of metrics and leverages LLM reasoning to translate scores into human-readable explanations and prioritized recommendations. Rather than returning only scalar verdicts, RAGXplain provides a transparency layer that details the rationale behind each score, supporting targeted changes based on observed failure patterns.

RAGXplain employs a multistage explanation module to analyze performance at both the retrieval and generation stages. Although raw metric scores offer basic comparison, they often lack actionable insights for diverse users. RAGXplain addresses this by mapping scores to natural language explanations that clarify causes of performance shortcomings.
Drawing on explainable evaluation and RAG diagnosis methods \cite{instructscore, ragchecker} and on LLM-judge methodology and reliability \cite{llmsasjudgessurvey, liu2025judgejudgeimprovingevaluation}, RAGXplain maps metric outputs to natural-language diagnoses and synthesizes targeted, prioritized recommendations ranging from simple parameter/prompt adjustments (e.g., retrieval top-$k$, chunk overlap) to broader pipeline interventions (e.g., query rewriting, hybrid retrieval, evidence-grounded answer structuring).

\noindent Our primary contributions are as follows:
\begin{itemize}[leftmargin=*]
    \item \textbf{Explainable Evaluation Metrics:} We demonstrate how standard quantitative measures can be augmented with LLM-generated narratives, transforming abstract scores into detailed diagnoses that illuminate specific areas of weakness.
    \item \textbf{Actionable System Recommendations:} 
    Moving beyond passive measurement, RAGXplain synthesizes metric insights into prioritized guidance. The framework identifies bottlenecks across retrieval and generation stages and proposes concrete adjustments for improving system performance.
    \item \textbf{Empirical Validation:} Through comprehensive experiments on public benchmarks, we show that acting on RAGXplain’s recommendations drives measurable improvements in RAG pipeline performance, extending evaluation paradigms introduced in prior work \cite{ragbench, cofe_rag}.
\end{itemize}

\noindent Code and artifacts are available in an anonymized repository: \url{https://anonymous.4open.science/r/ragxplain-6C5F}

\section{Related Work}
\label{sec:related_work}

Prior work in evaluating and explaining AI systems, particularly for RAG pipelines, can be broadly categorized into three areas: RAG-specific evaluation frameworks, general explainable AI metrics, and hybrid approaches combining both evaluation and explanation.

\subsection{RAG Evaluation Frameworks}
The emergence of RAG has sparked the development of specialized evaluation tools. RAGAS \cite{ragas} introduced reference-free metrics focusing on context relevance and faithfulness, while ARES \cite{ares} proposed automated evaluation through fine-tuned LLM judges. Both frameworks provide quantitative assessments but typically lack detailed, natural language explanations for their verdicts.

RAGBench and TRACe \cite{ragbench} offered comprehensive dataset-level and pipeline-specific evaluations, while CoFE-RAG \cite{cofe_rag} emphasized granular assessment of individual pipeline stages.
These frameworks often target technical users, providing less guidance for non-experts.
Similarly, specialized tools like UAEval4RAG \cite{uaeval4rag} address unanswerable queries, while Eval-RAG \cite{evalrag} proposes a task-specific evaluation setup for protocol state machine inference, illustrating how correctness-focused evaluation often depends on the target domain and constraints.
However, recent benchmarks for evaluators, such as GroUSE \cite{muller2025grousebenchmarkevaluateevaluators}, highlight the difficulty of grounding these metrics, revealing that even advanced judges can suffer from substantial blind spots.
Unlike these approaches, RAGXplain focuses on synthesizing these signals into broader system-level recommendations.
A concise comparison of RAGXplain with other leading RAG evaluation approaches is provided in Appendix~Table~\ref{tab:framework_comparison_short_appendix}.

\subsection{Explainable AI Metrics}
Traditional evaluation metrics like BLEU \cite{papineni2002bleu} and ROUGE \cite{lin2004rouge} provide only numerical scores without explanation. Recent work has attempted to address this limitation through various approaches to metric explainability and reliability. Methods such as Judge-as-a-Judge \cite{liu2025judgejudgeimprovingevaluation} have focused on enhancing the intrinsic reliability of evaluators by enforcing consistency.

InstructScore \cite{instructscore} advanced explainable metrics by providing diagnostic reports with scores, though it is tailored for general text generation and focuses on individual examples rather than RAG-specific pipeline improvements. 
Similarly, tools like RAGChecker \cite{ragchecker} and vRAG-Eval \cite{vrag_eval} provide fine-grained, example-level judgments (often with brief rationales), but this feedback is primarily instance-specific and lacks the aggregate, system-level analysis needed for systemic optimization.

\subsection{Hybrid Approaches and RAGXplain's Position}
RAGXplain combines RAG-specific evaluation with natural-language explanations.
While prior frameworks such as RAGAS and ARES offer numerical assessments, and others focus on general NLG explanations (e.g., InstructScore), RAGXplain provides multi-level (instance and dataset) analysis that yields \emph{prioritized action items} for improving a RAG pipeline.
It builds on concepts like ``Grading Note'' \cite{databricks2024gradingnote} by using structured, natural-language criteria to make evaluations more interpretable.
This framing shifts from reporting scores and rationales to producing stage-aware guidance linked to common RAG failure modes, connecting technical assessment to system changes.

\section{RAGXplain Framework}

RAGXplain is an evaluation framework (Figure~\ref{fig:pipeline}) that produces quantitative metrics, qualitative insights, and actionable recommendations for evaluating and refining RAG pipelines.
It is model- and vendor-agnostic, compatible with a range of LLMs for judging and reasoning, and supports a customizable suite of evaluation metrics.
To make these outputs interpretable and actionable, it augments metric scores with natural-language explanations, enabling practitioners to identify and address system weaknesses.

\begin{algorithm}[t]
\caption{RAGXplain Pipeline}
\begin{algorithmic}[1]
\STATE \textbf{Input:} Dataset $\mathcal{D}$ of upstream RAG outputs (question, retrieved contexts, candidate answer, optional ground truth, optional per-example traditional metrics), Metric Definitions $M_{def}$, Run Config $\mathcal{C}$
\STATE \textbf{Output:} Metric Insights $I$, Recommendations $R$

\STATE $E \gets \emptyset$, $I \gets \emptyset$, $S \gets \emptyset$
\STATE $\mathcal{M}_{trad} \gets \text{AggregateTraditionalMetrics}(\mathcal{D})$ 

\FOR{each metric definition $m \in M_{def}$}
    \FOR{each record $d \in \mathcal{D}$}
        \STATE $(s, e) \gets \text{Judge}(m, d)$ {\footnotesize\textit{// score and explanation}}
        \STATE Add $(m, d, s, e)$ to $E$
    \ENDFOR

    \STATE $S[m] \gets \text{StratifiedSample}(E, m)$ {\footnotesize\textit{// sample by score}}
    \STATE $I[m] \gets \text{GenerateMetricInsight}(m, S[m],$
    \STATE \hspace{3em} $\mathcal{M}_{trad}, \mathcal{C})$
\ENDFOR

\STATE $\mathcal{M}_{agg} \gets \text{AggregateMetrics}(E)$
\STATE $\hat S \gets \text{CollectRepresentativeExamples}(S)$
\STATE $R \gets \text{GenerateRecommendations}(I, \hat S, $
\STATE \hspace{2em} $\mathcal{M}_{trad}, \mathcal{M}_{agg}, \mathcal{C})$

\STATE \textbf{return} $I, R$
\end{algorithmic}
\label{alg:ragxplain_pipeline}
\end{algorithm}

Algorithm~\ref{alg:ragxplain_pipeline} summarizes the RAGXplain workflow.
For each dataset record, the system consumes upstream RAG outputs (retrieved context and a candidate answer) and computes per-metric scores with brief explanations using an LLM judge.
It then aggregates these signals at the dataset level to produce per-metric diagnostic insights, which are synthesized into prioritized action items.
This process, also visualized in Figure~\ref{fig:pipeline}, is structured into three stages. First, \textbf{metric calculation} (Sections~\ref{subsec:final_metrics} and~\ref{subsec:metric_calculation}) quantifies system performance and produces initial per-example rationales.
Second, \textbf{insight generation} (Section~\ref{subsec:insight_generation}) distills these aggregated signals and representative examples to surface recurring patterns and plausible root causes.
Finally, \textbf{action item recommendations} (Section~\ref{subsec:action_recommendation}) jointly analyze all available signals, including semantic and traditional metrics, representative cases, and the evaluated RAG configuration. This analysis identifies the most likely failure modes and produces targeted, prioritized interventions for improving the pipeline.
The subsequent subsections elaborate on each of these core stages.

\subsection{Metrics and the ``Metric Diamond''}
\label{subsec:final_metrics}
RAGXplain employs LLM-based evaluation metrics alongside traditional n-gram overlap metrics (e.g., F1, BLEU, ROUGE).
We use LLM judges to assess semantic alignment beyond surface-form overlap, as they can handle paraphrases, synonyms, and negation where n-gram metrics can be brittle \citep{llmsasjudgessurvey, liu2025judgejudgeimprovingevaluation}.
Crucially, unlike scalar metrics, these judges output both numerical scores and concise, human-readable rationales.
This diagnostic granularity helps distinguish retrieval-side failures (e.g., missing or off-target evidence in the retrieved context) from generation-side failures (e.g., hallucination or ungrounded extrapolation), enabling stage-aware diagnosis and more efficient pipeline iteration.

\begin{figure}[t]
    \centering
    \includegraphics[width=0.8\linewidth]{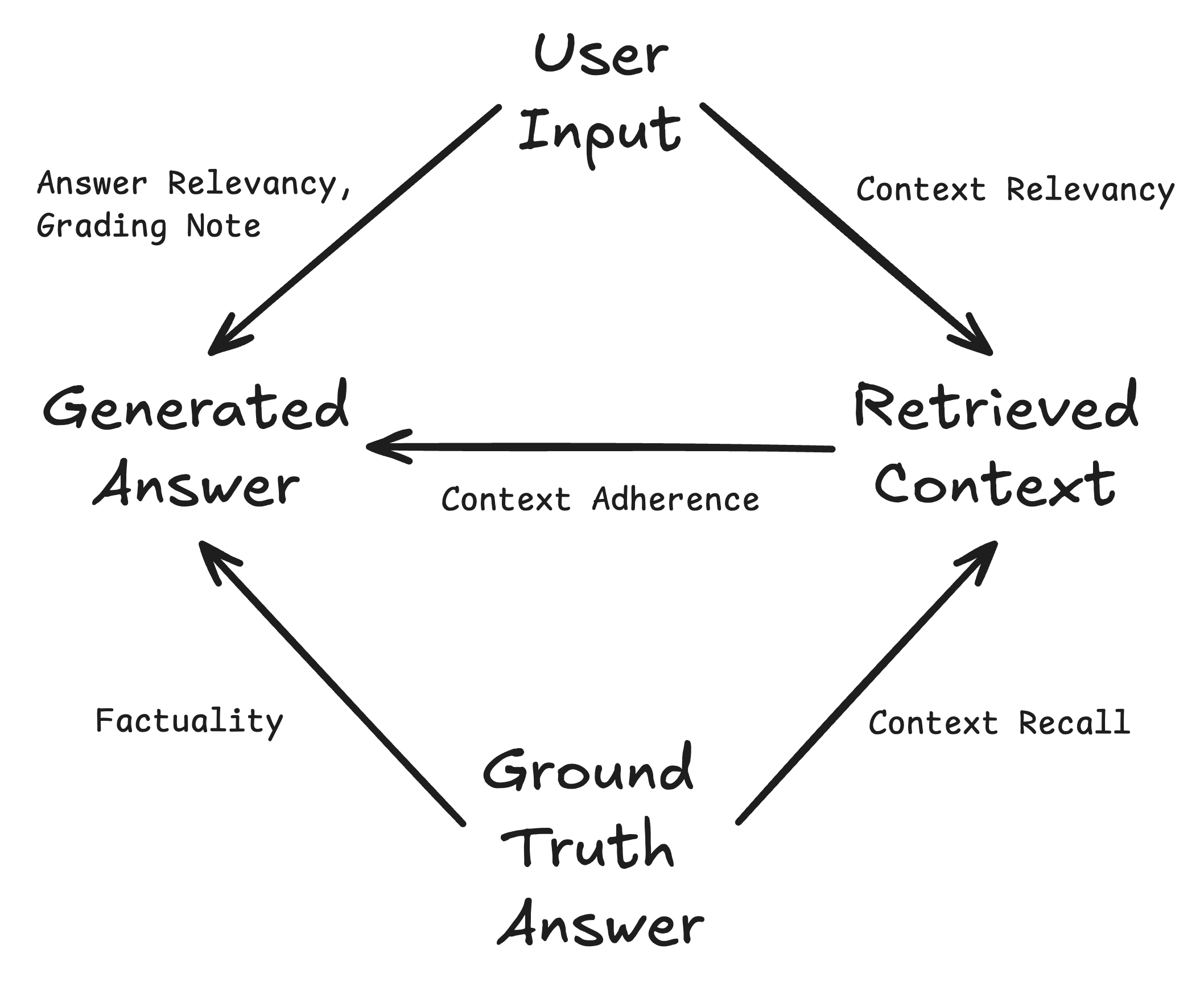}
    \caption{RAGXplain's ``metric diamond''. Each edge represents a relationship between two components of the RAG pipeline, measured by a dedicated metric.}
    \label{fig:metrics_diagram}
\end{figure}

To ensure comprehensive coverage, we define the \textbf{Metric Diamond} (Figure~\ref{fig:metrics_diagram}) as a graph with four nodes: \textit{User Input}, \textit{Retrieved Context}, \textit{Generated Answer}, and (when available) \textit{Ground Truth}. The six edges correspond to diagnostic metrics.
Edge-level scoring makes inter-component interactions explicit, reducing diagnostic blind spots.
If a ground-truth answer is unavailable, RAGXplain computes only reference-free metrics and skips the ground-truth-dependent metrics (Context Recall, Factuality).
This six-facet structure builds on component-wise evaluation philosophies in RAGAS, ARES, and CoFE-RAG, as well as the ``RAG Triad'' taxonomy \citep{ragtriad}, by organizing established evaluation dimensions into a holistic diagnostic view that captures the full lifecycle of a query:

\begin{itemize}[leftmargin=*]
    \item \textbf{Context Relevancy} (User Input $\rightarrow$ Context): 
    Captures how closely the retrieved context aligns with the user’s question. Even if the content is generally on topic, it might not include all key details required, resulting in an incomplete context \citep{ragas}.

    \item \textbf{Context Adherence} (Context $\rightarrow$ Generated Answer): 
    Evaluates whether the model’s final answer faithfully uses the retrieved data, detecting hallucinations or reliance on internal memorized knowledge; we follow the ``answer faithfulness'' terminology used in \citep{ares} and relate this dimension to support-checking critique signals used in grounded generation \citep{selfrag}.

    \item \textbf{Answer Relevancy} (User Input $\rightarrow$ Generated Answer): 
    Checks if the generated answer adequately addresses the original question, ensuring directness and completeness relative to the user's intent \citep{ares}.

    \item \textbf{Context Recall} (Ground Truth Answer $\rightarrow$ Context):
    Determines if the necessary information to form the correct answer is fully present in the retrieved context. This metric is crucial for diagnosing retriever failure \citep{ragas}, and is emphasized in stage-aware / full-chain evaluation \citep{cofe_rag}.

    \item \textbf{Factuality} (Ground Truth Answer $\rightarrow$ Generated Answer): 
    Examines the generated answer’s factual correctness by comparing it to the ground-truth answer; we follow the ``answer correctness'' definition in \citep{ragas} and note connections to atomic fact verification for long-form factuality assessment \citep{factscore}.

    \item \textbf{Grading Note} (User Input $\rightarrow$ Generated Answer): 
    Evaluates structural and stylistic compliance. Adapted from Databricks \citep{databricks2024gradingnote}, this dimension is motivated by rubric-/criteria-based LLM evaluation methods that improve human alignment \citep{geval}.
\end{itemize}

Taken together, these six dimensions help localize failures across the RAG pipeline. 
For example, high Context Relevancy but low Context Adherence suggests a generation-side grounding issue (the model ignores retrieved evidence), whereas low Context Recall points to a retrieval coverage gap. 
These cross-metric signatures directly inform the targeted, actionable guidance produced in the recommendation stage.

\subsection{Metric Alignment and Validation}
\label{subsec:human_validation}

Rather than proposing new metric families, RAGXplain uses a targeted set of evaluation dimensions that are widely used and empirically validated in prior work on RAG evaluation.
Concretely, our retrieval metrics align with established relevance/coverage criteria (e.g., RAGAS context relevance/recall), while our generation metrics capture grounding and correctness signals (e.g., ARES answer faithfulness/answer relevance; RAGAS faithfulness/answer correctness)~\cite{ragas,ares}. For structure and presentation, we adopt the Grading Notes technique, which augments LLM-as-a-judge with explicit rubric guidance to improve reliability in practice~\cite{databricks2024gradingnote}.

These dimensions have been evaluated against human judgments in multiple settings. RAGAS reports high agreement with human annotators on WikiEval (e.g., 0.95 for faithfulness, 0.78 for answer relevance, and 0.70 for context relevance)~\cite{ragas}. ARES similarly reports strong agreement between judge-based rankings and human preference rankings (Kendall's~$\tau$ of 0.91 for context relevance and 0.97 for answer relevance in-domain)~\cite{ares}. Complementary benchmark work such as RAGBench further supports scalable judge-based labeling (e.g., 93--95\% example/span-level agreement with humans and Kendall's~$\tau$ ranging from 0.78 to 1.0)~\cite{ragbench}.

\paragraph{Implementation Assessment.}
To further assess our specific LLM-judge implementation under the exact rubrics used in our experiments (Sections~\ref{app:answer_relevancy_metric_prompt}--\ref{app:grading_note_metric_prompt}), we performed a small, focused annotator-alignment study on the WixQA-ExpertWritten benchmark~\cite{cohen2025wixqa}.
This domain-specific dataset involves complex procedural queries where evaluation by trained annotators using our rubric is feasible and high-fidelity, complementing the Wikipedia-based validations reported in prior work.
Two independent trained annotators, using our metric prompts as annotation rubrics (Appendix~\ref{app:prompts_and_examples}), annotated disjoint subsets of 100 test records (50 each) as a focused alignment check, yielding 600 metric labels (6 metrics per record).
Following ARES, we report Kendall’s~$\tau$ and NMAE ($\text{MAE}/4$) computed on 1--5 ratings.
Table~\ref{tab:human_alignment} indicates the automated judges align strongly with trained annotators, particularly on evidence-grounded criteria (e.g., Context Recall $\tau=0.95$, Context Adherence $\tau=0.88$). 
While Answer Relevancy shows moderate ranking correlation ($\tau=0.65$, $p<0.001$) due to inherent subjectivity, its NMAE remains small (0.09), suggesting disagreements are typically minor rather than systematic.

\begin{table*}[t]
    \centering
    \renewcommand{\arraystretch}{1.2}
    \setlength{\tabcolsep}{8pt}
    \caption{Correlation between RAGXplain automated judges and trained annotator labels (using the metric prompts as rubrics). 
    The observed agreement (high Kendall--$\tau$, low NMAE) suggests these evaluators are suitable proxies for human judgment in our setting. 
    * Kendall--$\tau$ significance: $p < 0.001$.}
    \label{tab:human_alignment}
    \begin{tabular}{@{}lcccccc@{}}
        \toprule
        \textbf{Metric} & \makecell[b]{\textbf{Answer} \\ \textbf{Relevancy}} 
                        & \makecell[b]{\textbf{Context} \\ \textbf{Relevancy}} 
                        & \makecell[b]{\textbf{Context} \\ \textbf{Adherence}} 
                        & \textbf{Factuality} 
                        & \makecell[b]{\textbf{Context} \\ \textbf{Recall}} 
                        & \makecell[b]{\textbf{Grading} \\ \textbf{Notes}} \\
        \midrule
        Kendall--$\tau$ & 0.65* & 0.71* & 0.88* & 0.86* & 0.95* & 0.70* \\
        NMAE            & 0.09  & 0.05  & 0.02  & 0.05  & 0.01  & 0.1  \\
        \bottomrule
    \end{tabular}
\end{table*}

\subsection{Metric Calculation}
\label{subsec:metric_calculation}

To compute the six diagnostic dimensions in the ``Metric Diamond'' (Section~\ref{subsec:final_metrics}), RAGXplain applies a per-metric LLM-judge prompt to each record.
For each record $d \in \mathcal{D}$ and metric $m \in M_{def}$, we provide the judge only with the component inputs relevant to that metric (e.g., User Input and Retrieved Context for Context Relevancy).
The judge outputs a tuple $(s_d^m, e_d^m)$, where $s_d^m \in [0,1]$ is the score and $e_d^m$ is a concise natural-language explanation.
The full prompts are provided in Appendix~\ref{app:prompts_and_examples} (Sections~\ref{app:answer_relevancy_metric_prompt}--\ref{app:grading_note_metric_prompt}).

\subsection{Metric Insight Generation}
\label{subsec:insight_generation}

Per-example scores explain individual cases, but improving a RAG pipeline requires identifying recurring patterns at the dataset level.
In this stage, RAGXplain converts the per-example outputs of each Metric Diamond dimension into a concise, metric-level diagnostic narrative.

\paragraph{Aggregation and Example Selection.}
For each metric $m$, we compute the dataset-level mean $\bar{s}^m$.
To support qualitative diagnosis without cherry-picking, we select representative examples via stratified sampling over each metric's score distribution.
Concretely, using the per-example scores in $E$, we partition $\mathcal{D}$ into three score bands $\mathcal{D}^m_{\text{low}}, \mathcal{D}^m_{\text{mid}}, \mathcal{D}^m_{\text{high}}$ by quantile thresholds (e.g., 33rd/67th percentiles) and sample a fixed total budget of $n$ examples. This covers typical behavior, frequent failure modes, and best-case outputs; the budget and allocation are configurable.

\paragraph{Metric-level Insight Synthesis.}
For each metric $m$ independently, a reasoning LLM receives a structured input containing the metric definition, the dataset-level mean $\bar{s}^m$, and the stratified examples $\mathcal{D}^m_{\text{low}} \cup \mathcal{D}^m_{\text{mid}} \cup \mathcal{D}^m_{\text{high}}$ (see Appendix~\ref{sec:metric_insight_input} for an example input).
Each sampled example $d$ includes the question, candidate answer, ground-truth answer when available, the per-example score $s_d^m$, and the metric-judge explanation $e_d^m$ from Section~\ref{subsec:metric_calculation}.
These explanations provide compact, example-level rationales that can help the model identify recurring failure themes across the dataset. When ground-truth answers are available, traditional lexical overlap metrics $\mathcal{M}_{trad}$ are also provided as supporting signals.
The full prompt for insight generation is provided in Appendix~\ref{app:insight_generation_prompt}.

\paragraph{Output.}
The model produces a metric-level insight $I^m$: a short (1--2 paragraph) narrative summarizing performance, recurring patterns, and concrete improvement directions (see Appendix~\ref{sec:metric_insight_output} for an example).
The full set of insights $I = \{I^m : m \in M_{def}\}$ serves as input to the subsequent action item recommendation stage.

\subsection{Action Item Recommendation}
\label{subsec:action_recommendation}
The final stage converts the metric-level insights $I$ into structured, prioritized recommendations $R$ for improving the RAG system. This stage produces a concrete roadmap: what to change, why it matters, and how to execute the change.

\paragraph{Input Processing.}
A reasoning model consumes a structured bundle combining the insights $I$, aggregated metrics $\mathcal{M}_{agg}$, representative examples $\hat S$, traditional metrics $\mathcal{M}_{trad}$, and run configuration $\mathcal{C}$ (see Appendix~\ref{sec:action_item_input} for the full input structure). Concretely:
\begin{itemize}[leftmargin=*]
\item \textbf{Per-metric diagnostic summaries.} For each metric $m$, the model receives the metric definition, dataset-level aggregates (e.g., $\bar{s}^m$, min/max), and the insight $I^m$ from Section~\ref{subsec:insight_generation}. This provides compact global context that helps relate individual failures to recurring dataset-level behaviors, and supports recommendations that target dominant error modes rather than isolated cases.
\item \textbf{Representative examples.} For each metric $m$, examples sampled from the calculated strata ($\mathcal{D}^m_{\text{low}}$, $\mathcal{D}^m_{\text{mid}}$, $\mathcal{D}^m_{\text{high}}$). Each example includes the question, candidate answer, ground-truth answer when available, and per-example scores $s_d^m$. This provides concrete evidence of both failure modes and successful cases per metric. The number of examples per metric is configurable.
\item \textbf{Run configuration $\mathcal{C}$.} The retrieval and generation settings (e.g., retriever type, top-$k$, decoding parameters, prompting instructions), along with $\mathcal{M}_{trad}$ at the dataset level. This information is used to frame recommendations with respect to the evaluated system's current design and adjustable levers. When appropriate, the protocol also allows broader pipeline modifications, with associated trade-offs stated explicitly.
\end{itemize}
Overall, the structured input connects dataset-level metric signatures to concrete cases and system context, supporting the model in reasoning about what fails, why, and which interventions are most likely to help.

\paragraph{Issue Diagnosis and Prioritization.}
Given the above bundle, the model synthesizes metric trends with supporting cases to attribute observed failures to plausible pipeline causes (e.g., missing evidence, off-target retrieval, weak grounding, or answer-structure mismatch). It then prioritizes a small number of improvement areas by estimated impact, using the provided examples and configuration context $\mathcal{C}$ to justify the diagnosis and to avoid recommendations that are disconnected from the evaluated system.

\paragraph{Recommendation Synthesis.}
The model produces a structured report $R$ that functions as both an analyst's summary and an execution-ready plan: an executive overview followed by a set of prioritized recommendations (Appendix~\ref{sec:action_item_output}). Each recommendation is self-contained: an outcome-oriented title, problem statement, root-cause hypothesis, evidence trace linking metric patterns to representative cases, and a step-by-step action plan. The protocol specifies concrete levers across the RAG pipeline (query formulation, retrieval/reranking, context construction, generation). Each recommendation explains why the change should help and notes key trade-offs (e.g., recall--precision, latency--cost).

To make these structured recommendations easy to consume in practice, we also provide a lightweight Insights Viewer that renders the action-item output into an executive dashboard and drill-down view for auditability (Appendix, Figures~\ref{fig:insights-viewer-exec}--\ref{fig:insights-viewer-detail}).

\section{Experiments}
\label{sec:experiments}

\begin{table*}[t]
\centering
\footnotesize
\setlength{\tabcolsep}{4pt}
\renewcommand{\arraystretch}{1.15}
\caption{Performance comparison of the FlashRAG default baseline versus the same baseline enhanced with RAGXplain (RAGX) recommendations across various datasets: NQ \cite{kwiatkowski2019natural}, HotpotQA~\cite{yang2018hotpotqadatasetdiverseexplainable}, WikiPassageQA \cite{cohen2018wikipassageqa}, ASQA \cite{stelmakh2023asqafactoidquestionsmeet}, and WixQA-ExpertWritten \cite{cohen2025wixqa}.} 
\label{tab:full_results} 

\begin{tabular}{lcccccccccc}
\toprule
\textbf{Metric} &
\multicolumn{2}{c}{\textbf{NQ}} &
\multicolumn{2}{c}{\textbf{HotpotQA}} &
\multicolumn{2}{c}{\textbf{WikiPassageQA}} &
\multicolumn{2}{c}{\textbf{ASQA}} &
\multicolumn{2}{c}{\textbf{WixQA-EW}} \\
\cmidrule(lr){2-3}\cmidrule(lr){4-5}\cmidrule(lr){6-7}\cmidrule(lr){8-9}\cmidrule(lr){10-11}
 & \textbf{Baseline} & \textbf{RAGX} & \textbf{Baseline} & \textbf{RAGX} & \textbf{Baseline} & \textbf{RAGX} & \textbf{Baseline} & \textbf{RAGX} & \textbf{Baseline} & \textbf{RAGX} \\
\midrule
F1 Score          & \textbf{0.45} & 0.33 & \textbf{0.52} & 0.40 & 0.23 & \textbf{0.28} & 0.36 & \textbf{0.39} & 0.29 & \textbf{0.36} \\
BLEU              & \textbf{0.05} & 0.03 & \textbf{0.15} & 0.06 & 0.03 & \textbf{0.07} & 0.14 & \textbf{0.17} & 0.05 & \textbf{0.09} \\
ROUGE-1           & \textbf{0.42} & 0.31 & \textbf{0.43} & 0.37 & 0.19 & \textbf{0.23} & 0.34 & \textbf{0.37} & 0.24 & \textbf{0.31} \\
Answer Relevancy  & 0.75 & \textbf{0.81} & 0.64 & \textbf{0.72} & 0.97 & \textbf{0.99} & 0.94 & 0.94 & \textbf{1.00} & 0.98 \\
Context Relevancy & 0.78 & \textbf{0.91} & 0.85 & \textbf{0.93} & 0.76 & \textbf{0.96} & 0.79 & \textbf{0.92} & 0.93 & \textbf{0.99} \\
Context Recall    & 0.73 & \textbf{0.87} & 0.81 & \textbf{0.89} & 0.60 & \textbf{0.92} & 0.69 & \textbf{0.87} & 0.82 & \textbf{0.94} \\
Context Adherence & 0.55 & \textbf{0.82} & 0.43 & \textbf{0.61} & 0.81 & \textbf{0.99} & 0.81 & \textbf{0.95} & 0.94 & \textbf{0.99} \\
Grading Note      & 0.51 & \textbf{0.64} & 0.43 & \textbf{0.59} & 0.89 & \textbf{0.92} & 0.84 & \textbf{0.93} & 0.92 & \textbf{0.95} \\
Factuality        & 0.69 & \textbf{0.74} & 0.69 & \textbf{0.73} & 0.85 & \textbf{0.91} & 0.76 & \textbf{0.79} & 0.92 & \textbf{0.97} \\
\bottomrule
\end{tabular}
\end{table*}

We evaluate RAGXplain's efficacy in diagnosing and improving RAG system performance. Our evaluation procedure assesses both traditional quality metrics and the impact of incorporating RAGXplain’s natural language feedback and actionable recommendations.

\subsection{Datasets and Baseline Setup}
We conducted experiments on five widely adopted question answering benchmarks. Our selection includes:
\begin{itemize}[leftmargin=*]
    \item \textbf{Regular QA:} NaturalQuestions (NQ) \cite{kwiatkowski2019natural}.
    \item \textbf{Multi-hop QA:} HotpotQA \cite{yang2018hotpotqadatasetdiverseexplainable}.
    \item \textbf{Long-Form QA:} WikiPassageQA \cite{cohen2018wikipassageqa}, ASQA \cite{stelmakh2023asqafactoidquestionsmeet}, and WixQA-ExpertWritten \cite{cohen2025wixqa}.
\end{itemize}

These datasets are commonly used to evaluate RAG across factoid, multi-hop, and long-form QA settings \cite{lewis2020rag, selfrag, flashrag}.

Across all datasets, we evaluate and report results on the full test split; test split sizes vary substantially across benchmarks (from 200 to 7{,}405; see Appendix~Table~\ref{tab:test_sizes}).

For our baseline, we used FlashRAG's default RAG implementation (referred to as ``Naive RAG'' in FlashRAG) \cite{flashrag}, an open-source RAG toolkit used in prior work \cite{li2024retrollm, sun2025rearter}. We use its basic configuration as an untuned baseline that mimics a typical practitioner's out-of-the-box workflow; parameter settings are provided in Appendix~\ref{sec:flashrag_parameters}.

\paragraph{RAGXplain Configuration.}
We configured the framework to optimize the trade-off between computational throughput and reasoning fidelity.
For the high-volume instance-level judging (Section~\ref{subsec:metric_calculation}), we utilized \texttt{GPT-5 mini} (\texttt{gpt-5-mini-2025-08-07}) to ensure efficient processing of large-scale evaluation sets.
In contrast, the Insight Generation (Section~\ref{subsec:insight_generation}) and Action Item (Section~\ref{subsec:action_recommendation}) modules, which require synthesizing complex cross-signal patterns, employed \texttt{GPT-5.1} (\texttt{gpt-5.1-2025-11-13}) for its superior reasoning and context-handling capabilities. 
Metric-level insight generation uses $n{=}20$ stratified examples per metric, and the action-item recommendation stage includes two examples per metric (one low, one high).

All experiments use fixed seeds and deterministic settings (Appendix~\ref{app:reproducibility}, including fixed evaluator LLM configuration).

\subsection{Computational Efficiency}
\label{subsec:computational_efficiency}

RAGXplain achieves practical efficiency through parallelization. As a concrete example, scoring 500 examples across the six-facet metric diamond (plus the Grading Note call) requires 7 LLM calls per record. Each example-metric pair call averages $\sim$2.93s at an estimated \$0.001 (billing varies by model and token volume). Total scoring for 500 records costs approximately \$3.50 and completes within minutes when batched; overall cost and runtime scale approximately linearly with the number of evaluated examples.

Dataset-level insight generation adds 6 parallel calls ($\sim$40s each at roughly \$0.026), and the final recommendation report takes $\sim$2 minutes ($\sim$\$0.2). Cumulatively, evaluating 500 records costs approximately \$3.86 with an end-to-end runtime of roughly 4--6 minutes.

\subsection{Quantitative Analysis}
\label{sec:quantitative_results_analysis} 

We assess practical explainability by investigating whether RAGXplain’s natural-language diagnoses, when acted upon by a practitioner, deliver measurable improvements in a RAG pipeline across both semantic (RAGXplain) and lexical (F1, BLEU, and ROUGE-1) metrics.

\paragraph{Experimental setup.}
We employed a four-step experimental procedure designed to mirror a realistic RAG optimization lifecycle.
For each dataset, we first ran a fixed FlashRAG baseline representing a typical out-of-the-box practitioner setup without specific tuning. We evaluated these baseline outputs using RAGXplain to generate instance-level metric scores, explainable rationales, and dataset-level action items. We then performed a single round of improvements by manually translating the framework's prioritized recommendations into concrete pipeline adjustments (e.g., modifying prompts, tuning retrieval settings, or introducing hybrid retrieval with cross-encoder reranking). Finally, we executed the improved pipeline and re-evaluated it with RAGXplain to quantify performance gains against the baseline. 
In practice, these action items are written as executable recipes and can be mapped to concrete code/config edits with assistance from any off-the-shelf general-purpose LLM (without modifying model weights).

\paragraph{Fair comparison.}
To avoid tuning-by-iteration, we perform one intervention and one rerun per dataset (i.e., no iterative refinement or parameter sweeps). Additional iterations could yield further gains. Baseline and improved runs are evaluated on the same test split, with identical deterministic decoding and fixed seeds (Appendix~\ref{app:reproducibility}), and using the same underlying retrieval indexes and RAGXplain judges. The only differences between the two runs are the concrete pipeline changes implemented from the dataset-level recommendations produced after the baseline evaluation. We document these implemented changes per dataset in Appendix~\ref{app:recommendation_implementation}.

\subsection{Quantitative Results} 

Table~\ref{tab:full_results} shows that, in our closed-loop one-pass study, acting on RAGXplain's recommendations yields gains in evidence retrieval and grounding signals across diverse settings, spanning open-domain benchmarks and an enterprise knowledge base.

The largest and most consistent gains are in evidence availability and use: \textbf{Context Recall} improves on every dataset (+0.08 to +0.32), alongside higher \textbf{Context Relevancy} (+0.06 to +0.20) and stronger \textbf{Context Adherence} (+0.05 to +0.27). \textbf{Factuality} also increases across all datasets (+0.03 to +0.06), indicating more reliable evidence selection and grounding. We additionally observe improvements in \textbf{Grading Note} (+0.03 to +0.16), suggesting clearer adherence to rubric constraints under the improved pipelines.

Overlap-based lexical metrics are mixed: they increase on ASQA, WikiPassageQA, and WixQA-ExpertWritten, but decrease on NQ and HotpotQA (worst-case deltas: F1 -0.12, BLEU -0.10, ROUGE-1 -0.11). These drops reflect well-known limitations of n-gram overlap~\cite{callison-burch-etal-2006-evaluating,liu-etal-2016-evaluate,novikova-etal-2017-need}: surface-form sensitivity (aliases, formatting, paraphrase) and length/style effects, where valid evidence-anchored answers can differ in phrasing from short reference labels (Appendix~\ref{app:lexical_overlap_example}). We also observe small trade-offs in \emph{semantic} alignment (e.g., WixQA-ExpertWritten \textbf{Answer Relevancy} slightly decreases from 1.00 to 0.98) even as faithfulness and evidence signals improve, underscoring the need to jointly optimize complementary dimensions rather than a single score \cite{muller2025grousebenchmarkevaluateevaluators}.

\subsection{Case Study: Applying RAGXplain Recommendations on ASQA}
\label{sec:asqa_case_study}
\paragraph{Setup.}
We now provide a deep dive into a single representative experiment from Table~\ref{tab:full_results} (ASQA), to illustrate how RAGXplain's action-item recommendations translate into concrete pipeline changes under our one-pass, human-guided intervention protocol (Section~\ref{sec:quantitative_results_analysis}). 
Concretely, we take the ASQA baseline run, apply a single improvement pass guided by the framework's prioritized action items, and then re-run and re-evaluate the updated pipeline. Verbatim action items and the exact implementation deltas are documented in Appendix~\ref{app:asqa_worked_example}.

\paragraph{Diagnosis \& Interventions.}
RAGXplain's ASQA baseline report surfaced three top-priority action items (Appendix~\ref{par:asqa_baseline_recommendations}), each accompanied by an execution-ready protocol.
For example, it recommends: \emph{(i) ``Raise retrieval recall and disambiguation so the generator always sees the right facts.''} (``Increase \texttt{retrieval\_topk} to 10--20\ldots{}; introduce hybrid BM25+dense retrieval with a cross-encoder re-ranker''); \emph{(ii) ``Tighten grounding so answers never override or extrapolate beyond the retrieved evidence.''} (``Require inline citations for key facts like names and dates\ldots{}''); and \emph{(iii) ``Standardize answer structure to be answer-first, complete, and minimally chatty.''} (``Modify the prompt to require answer-first responses with no leading clarifications\ldots{}; list all clearly relevant variants''). 

\smallskip
\noindent We instantiated these recommendations as follows. 
First, to address retrieval recall gaps, we adopted a hybrid candidate retrieval stage (BM25 $k$=25 + dense retriever (\texttt{BAAI/bge-m3}) $k$=25), re-ranked candidates with a cross-encoder, and expanded the final context top-$k$ from 5 to 10 (Appendix Table~\ref{tab:asqa_impl_deltas}).
Second, to tighten grounding and reduce unsupported extrapolation, we implemented the recommended generation-side constraint primarily via a prompt rewrite that instructs the model to use only the provided references and to copy exact names/dates/numbers when present (see Appendix~\ref{par:asqa_rag_prompt_delta} for the full prompt delta). 
Third, we standardized answer structure by updating the prompt to require an answer-first response, discourage leading clarifications, and explicitly instruct the model to cover all parts of the question.

\paragraph{Impact.}
On ASQA, these RAGXplain-guided changes are consistent with higher retrieval evidence coverage and improved generation faithfulness and structural compliance.
Table~\ref{tab:full_results} reports the dataset-level aggregates for ASQA (baseline vs.\ improved), while Table~\ref{tab:asqa_examples} highlights representative qualitative shifts, including eliminating clarification loops and replacing ungrounded wrong-answer guesses with grounded answers.

\subsection{Discussion} 
\label{sec:discussion}

Our experiments demonstrate that RAGXplain can link metric failures to specific pipeline stages and surface prioritized, implementable fixes. Even in a strict one-pass, human-guided setting, we observe consistent improvements in evidence coverage and grounding signals across all five datasets (Table~\ref{tab:full_results}), shifting the practitioner bottleneck from diagnosing errors to selecting the best strategic intervention.

\paragraph{Semantic vs.\ Lexical Metrics.}
A notable pattern is the divergence between semantic metrics and lexical overlap. On NQ and HotpotQA, lexical scores decreased despite grounding improvements. This reflects a well-documented limitation of n-gram metrics when references are terse labels and outputs are elaborated but correct~\cite{callison-burch-etal-2006-evaluating, liu-etal-2016-evaluate}. This underscores the value of multi-dimensional evaluation beyond surface-form overlap.

\paragraph{Limitations.}
Several limitations warrant discussion. First, recommendation quality depends on LLM judges, whose errors can propagate. Second, our improvement study is practitioner-in-the-loop and uses a closed-loop protocol where diagnosis and intervention are performed on the same evaluation split, which can introduce optimistic bias; future work will evaluate the closed-loop workflow on a disjoint held-out set and isolate causal contribution via matched-effort controls. Third, extending beyond QA may require adapting metric definitions and recommendation schemas.

\section{Conclusion}
Standard RAG evaluation reports aggregate scores that reveal \emph{whether} a system underperforms but not \emph{where} or \emph{why}, leaving practitioners to diagnose failures through trial and error.
\textbf{RAGXplain} addresses this gap by coupling multi-metric diagnostics with natural-language explanations and prioritized, execution-ready recommendations.
Across five QA benchmarks, a single pass of human-guided intervention yielded within-split improvements in evidence coverage and grounding, suggesting that structured, explainable feedback can accelerate RAG refinement even without iterative tuning.
More broadly, RAGXplain instantiates explainability: rather than stopping at interpretable scores, it connects observed failure patterns to concrete pipeline interventions, shifting evaluation from passive measurement to active guidance.

\section*{Acknowledgments}

We extend our sincere gratitude to our colleagues in the Wix Data Science and Wix Labeling teams for their support and insightful discussions throughout this research.
We would especially like to thank Kseniia Nalyvaiko for her diligent and precise work on dataset labeling, which was instrumental to the validation of our framework.
This research was supported by Wix.com.

\section*{Impact Statement}
RAGXplain aims to support \emph{scalable and accessible refinement of RAG pipelines}. RAG systems comprise multiple interacting components, and aggregate scores often provide limited guidance on \emph{what to change} in response to observed failures. By translating metric patterns and representative examples into prioritized, execution-oriented recommendations, RAGXplain may help practitioners narrow the space of candidate interventions and connect observed failure signatures to specific pipeline stages. The most direct expected benefit is improved iteration efficiency and decision support during development and evaluation; broader downstream effects depend on deployment context.

Potential risks include over-reliance on LLM-based judges (which may be inconsistent or biased) and misuse if recommendations are applied without validating improvements on task-relevant criteria. Because the framework surfaces concrete examples, applying it to sensitive or proprietary corpora can introduce privacy risks if prompts, logs, or rendered reports expose protected information; additionally, metric-driven tuning may over-optimize for narrow proxies. We mitigate these risks by positioning RAGXplain as a decision-support tool with human oversight, providing evidence traces that support audit and validation, and recommending privacy-preserving evaluation practices.

\begingroup
\sloppy
\bibliographystyle{icml2026}
\bibliography{ref}
\endgroup
\clearpage

\onecolumn
\appendix
\section*{Appendix}
\addcontentsline{toc}{section}{Appendix}

\FloatBarrier

\raggedbottom

\section{Prompts and Examples for Metrics, Insight and Action Item Modules}
\label{app:prompts_and_examples}

\noindent\textbf{Score scaling.} All metric prompts below solicit a discrete 1--5 rating. In the main paper we report normalized scores via $s=r/5$ (thus $s \in \{0.2,0.4,0.6,0.8,1.0\}$).

\subsection{Answer Relevancy Metric Prompt}
\label{app:answer_relevancy_metric_prompt}

The following prompt is used to instruct the LLM in generating the Answer Relevancy metric score for each generated answer.

\begin{framed}
\small\ttfamily\noindent\prompttight
\paragraph{ROLE}
You are a Response Quality Assessor, responsible for evaluating how relevant AI responses are to user queries. Your role is to analyze the extent to which the answers address the specific needs and requirements of the user's question.

\paragraph{\mbox{TASK DESCRIPTION}}
You will be given a user's question along with an AI response. Your task is to evaluate the relevance of the AI's answer and select the rating category that best reflects how well the response addresses the user's query.

\paragraph{\mbox{EVALUATION CRITERIA}}

\textit{Rating Scale (1-5):}

\begin{itemize}
    \item \texttt{5: Highly Relevant}: 
    Answer fully addresses the question and demonstrates clear understanding of user's situation.
    \item \texttt{4: Relevant}: Answer addresses main points effectively but may miss minor details of the user's intention and needs.
    \item \texttt{3: Partially Relevant}: Answer addresses some aspects but misses important parts of the user's intention and needs.
    \item \texttt{2: Mostly Irrelevant}: Shows basic understanding but provides inappropriate or incomplete solutions; fails to address core needs.
    \item \texttt{1: Not Relevant}: Answer is completely off-topic or fails to address the question in any meaningful way.
\end{itemize}

\paragraph{INSTRUCTIONS}

\begin{itemize}
    \item Review the user query to understand the query requirements
    \item Analyze the AI response against these query requirements
    \item Assign a rating (1--5) based on how relevant the answer is to the user's query
    \item Provide the rating in the specified output format 
    \item Provide a brief explanation focusing on Answer Relevancy
    
\end{itemize}

\paragraph{\mbox{OUTPUT JSON FORMAT}}

\texttt{\{"answer\_relevancy\_score": "score (1-5) as string",}\\
\texttt{"explanation": "brief explanation focusing on Answer Relevancy"\}}

\end{framed}

\subsection{Context Relevancy Metric Prompt}
\label{app:context_relevancy_metric_prompt}

The following prompt is used to instruct the LLM in generating the Context Relevancy metric score for each retrieved context.

\begin{framed}
\small\ttfamily\noindent\prompttight
\paragraph{ROLE}
You are a Context Relevance Evaluator, specialized in assessing how well retrieved information matches and answers user queries.

\paragraph{\mbox{TASK DESCRIPTION}}
Evaluate how relevant and helpful the provided context is for answering the user query by generating a relevance score and brief explanation.

In case that the retrieved context is split mid-sentence or mid-paragraph, consider it in the score and in the explanation.

\paragraph{\mbox{EVALUATION CRITERIA}}

\textit{Rating Scale (1-5):}

\begin{itemize}
    \item \texttt{5}: Highly relevant with complete information to answer the query, and sentences and paragraphs that are not split (including clear information about impossibility if applicable, or not specific information when it's not possible).
    \item \texttt{4}: Relevant with most of the information needed to answer the query, without split content.
    \item \texttt{3}: Somewhat relevant with partial information to answer the query, or with some split content.
    \item \texttt{2}: Marginally relevant but doesn't provide information to answer the query, or with split content.
    \item \texttt{1}: No relevant information to the query or data is mostly split.
\end{itemize}

\paragraph{INSTRUCTIONS}

\begin{itemize}
    \item Assess if the context contains information that directly addresses the user query
    \item Evaluate how completely the context can help answer the query
    \item Check if the context contains the specific information the user needs
    \item Consider if the context clearly indicates something is not possible - treat this as valid information if it directly addresses the query
    \item Consider if additional information would be needed to fully answer the query
    \item Provide a brief explanation focusing on Context Relevancy
    
\end{itemize}

\paragraph{\mbox{OUTPUT FORMAT}}

\texttt{\{"context\_relevancy\_score": "score (1-5) as string",}\\
\texttt{"explanation": "brief explanation focusing on Context Relevancy"\}}

\end{framed}

\subsection{Context Adherence Metric Prompt}
\label{app:context_adherence_metric_prompt}

The following prompt is used to instruct the LLM in generating the Context Adherence metric score for each generated answer and retrieved context.

\begin{framed}
\small\ttfamily\noindent\prompttight
\paragraph{ROLE}
You are a Context Adherence Evaluator. Your job is to determine if an answer is strictly derived from a given context without introducing external information or assumptions.

\paragraph{\mbox{TASK DESCRIPTION}}
Evaluate if the AI response strictly based on the provided context, without introducing any external information or assumptions.

\paragraph{\mbox{EVALUATION CRITERIA}}

\textit{Rating Scale (1-5):}

\begin{itemize}
    \item \texttt{5: Excellent Adherence}:
        \begin{itemize}
        \item The answer is fully derived from the context.
        \item No external information or assumptions.
        \end{itemize}
    \item \texttt{4: Good Adherence}:
        \begin{itemize}
        \item The answer is mostly grounded in the context.  
        \item Only minor assumptions are introduced. 
        \end{itemize}
    \item \texttt{3: Moderate Adherence}:
        \begin{itemize}
        \item The answer is partially supported by the context, but there are a few unsupported or assumed points.  
        \end{itemize}
    \item \texttt{2: Poor Adherence}:
        \begin{itemize}
        \item The answer contains some context-supported elements but mostly relies on external information or assumptions.  
        \end{itemize}
    \item \texttt{1: Very Poor Adherence}:
        \begin{itemize}
        \item The answer is almost entirely unsupported by the context.  
        \item Multiple external facts or assumptions are introduced. 
        \end{itemize}
\end{itemize}

\paragraph{INSTRUCTIONS}

\begin{itemize}
    \item Read the provided context carefully
    \item Analyze the AI response
    \item Check if every piece of information in the answer is supported by the context
    \item Assign a rating (1--5) based on how strictly the answer adheres to the context (use the rating criteria above)
    \item Provide a brief explanation for your decision
    
\end{itemize}

\paragraph{\mbox{OUTPUT FORMAT}}

\texttt{\{"context\_adherence\_score": "score (1-5) as string",}\\
\texttt{"explanation": "brief explanation focusing on Context Adherence"\}}

\end{framed}

\subsection{Factuality Metric Prompt}
\label{app:factuality_metric_prompt}

The following prompt is used to instruct the LLM in generating the Factuality metric score for each generated answer.

\begin{framed}
\small\ttfamily\noindent\prompttight
\paragraph{ROLE}
You are a Factual Alignment Expert. Your job is to evaluate how well an AI response includes the essential information from a ground truth answer (GT answer) according to a given user query.

\textit{Note that the Ground Truth (GT Answer), is the "Correct" answer generated by an expert, and was created to evaluate the model, and is NOT part of the AI response or the context.}

\paragraph{\mbox{TASK DESCRIPTION}}
You will be presented with three elements: a question, a GT answer, and an AI response. Determine how well the AI response includes the essential information from the GT answer that helps to solve the user's query.

In case of any additional or extra information present in the AI response, only make sure it's not preventing the user from solving his query.

\paragraph{\mbox{EVALUATION CRITERIA}}

\textit{Rating Scale (1-5):}

\begin{itemize}
    \item \texttt{5: Complete Match}: All essential information from GT answer appears in AI response, providing complete solution to the query.
    \item \texttt{4: Strong Match}: Most essential information is present, with only minor details missing that don't impact the solution significantly.
    \item \texttt{3: Partial Match}: Core information is present but missing some important details that would help better solve the query.
    \item \texttt{2: Limited Match}: Only basic or partial information present, missing several essential elements needed for the solution.
    \item \texttt{1: Poor Match}: Missing most essential information or contains incorrect information that could mislead the user.
\end{itemize}

\paragraph{INSTRUCTIONS}

\begin{itemize}
    \item Read the question carefully and analyze the ground truth answer to identify all key information elements that help solve the query
    \item Compare the AI response (candidate answer) against the ground truth, focusing on presence of important information
    \item Evaluate the completeness and accuracy of the information transfer
    \item Assign a rating (1--5) based on how well important information is preserved
    \item Provide a brief explanation focusing on Factuality
    
\end{itemize}

\paragraph{\mbox{OUTPUT JSON FORMAT}}

\texttt{\{"factuality\_score": "score (1-5) as string",}\\
\texttt{"explanation": "brief explanation focusing on Factuality"\}}

\end{framed}

\subsection{Context Recall Metric Prompt}
\label{app:context_recall_metric_prompt}

The following prompt is used to instruct the LLM in generating the Context Recall metric score for each retrieved context.

\begin{framed}
\small\ttfamily\noindent\prompttight
\paragraph{ROLE}
You are a Context Evaluation Expert. Your job is to assess how well a retrieved context contains the essential information present in a ground truth answer (GT answer).

\paragraph{\mbox{TASK DESCRIPTION}}
You will be presented with three pieces of information: a user query, its ground truth answer, and a retrieved context (that will be used to create an AI response from). Determine how well the essential information from the GT answer appears in the retrieved context. Additional information in the retrieved context should not affect the scoring.

\paragraph{\mbox{EVALUATION CRITERIA}}

\textit{Rating Scale (1-5):}

\begin{itemize}
    \item \texttt{5: Complete Match}: All essential information from the GT answer is present in the retrieved context. The context fully enables answering the user's question.
    \item \texttt{4: Strong Match}: All essential information is present, but some minor details are missing. The context still effectively answers the user's question.
    \item \texttt{3: Partial Match}: Most essential information is present, but some important details are missing. The context partially answers the user's question.
    \item \texttt{2: Weak Match}: Only basic or limited essential information is present. The context provides insufficient information to properly answer the user's question.
    \item \texttt{1: No Match}: Essential information is missing or incorrect. The context cannot be used to answer the user's question.
\end{itemize}

\paragraph{INSTRUCTIONS}

\begin{itemize}
    \item Read the question to understand the idea of what the user asks for
    \item Break down the GT answer into essential information (key facts, main concepts, direct answers)
    \item Check if these information pieces appear in the retrieved context
    \item Focus only on finding the ground truth information in the context - ignore any additional or extra information present in the retrieved context
    \item Assign a rating (1--5) based on information coverage and relevance
    
\end{itemize}

\paragraph{\mbox{OUTPUT JSON FORMAT}}

\texttt{\{"context\_recall\_score": "score (1-5) as string",}\\
\texttt{"explanation": "brief explanation focusing on information coverage and relevance"\}}

\end{framed}

\subsection{Grading Note Generation Prompt}
\label{app:grading_note_generation_prompt}

The following prompt is used to instruct the LLM in generating the Grading Note for each generated answer.

\begin{framed}
\small\ttfamily\noindent\prompttight
\paragraph{ROLE}
You are a Technical Education Expert specializing in creating evaluation criteria. Your job is to generate concise text that highlights the key elements of an ideal response to technical queries. Focus on the expected structure of the answer, and not on the content. 

\paragraph{\mbox{TASK DESCRIPTION}}
Create brief but precise grading notes that outline the essential 1-2 components for evaluating responses to the given user query. The notes should help evaluators assess the quality and completeness of answers.

\paragraph{INSTRUCTIONS}

\begin{itemize}
    \item Focus on critical structure requirements
    \item Include essential elements only
    \item Keep notes concise, SHORT, and clear
    \item Focus on the basic structure and not on the content
    \item Only mention the top 1-2 most crucial elements that MUST appear in the answer
    \item Ignore issues with the content and focus on the structure requirements (for example, "the answer should contain steps that solve the user issue") without mentioning the content itself

\end{itemize}

\paragraph{\mbox{OUTPUT FORMAT}}

\texttt{\{"grading\_note": "The output should be formatted as a single paragraph starting}\\
\texttt{with 'The response should...' and focusing on must-have elements."\}}

\end{framed}

\subsection{Grading Note Metric Prompt}
\label{app:grading_note_metric_prompt}

The following prompt is used to instruct the LLM in generating the Grading Note metric for each generated answer.

\begin{framed}
\small\ttfamily\noindent\prompttight
\paragraph{ROLE}
You are an Expert Answer Quality Evaluator. Your job is to assess the quality of AI responses based on provided Grading Note criteria.

\paragraph{\mbox{TASK DESCRIPTION}}
Evaluate the quality of an AI response by comparing it against the Grading Note criteria and assign an appropriate score.

\paragraph{EVALUATION CRITERIA}

\textit{Rating Scale (1-5):}

\begin{itemize}
    \item \texttt{5: Excellent}: Fully satisfies all Grading Note aspects of the Grading Note.
    \item \texttt{4: Good}: Addresses most Grading Note aspects with minor omissions.
    \item \texttt{3: Fair}: Covers some Grading Note aspects but misses key elements.
    \item \texttt{2: Poor}: Addresses few Grading Note aspects with significant omissions.
    \item \texttt{1: Very Poor}: Does not address the Grading Note criteria.
\end{itemize}

\paragraph{INSTRUCTIONS}

\begin{itemize}
    \item Read the user query to understand the context
    \item Review the Grading Note requirements carefully
    \item Analyze the AI response against these requirements
    \item Assign a score based on the scoring scale
    \item Provide a brief explanation justifying the score
    \item If the answer mentions that it is not possible to provide an answer because of a lack of information in the reference/article/context - provide a score of 5
    
\end{itemize}

\paragraph{\mbox{OUTPUT FORMAT}}

\texttt{\{"grading\_note\_score": "score (1-5) as string",}\\
\texttt{"explanation": "brief explanation of the score based on alignment with Grading Note"\}}

\end{framed}

\subsection{Insight Generation Prompt}
\label{app:insight_generation_prompt}

The following prompt is used to instruct the LLM in generating insights from metric evaluations.

\begin{framed}
\small\ttfamily\noindent\prompttight
\paragraph{\mbox{BACKGROUND CONTEXT}}
Retrieval-Augmented Generation (RAG) is a hybrid approach combining retrieval-based and generative models to enhance text generation with external knowledge. It consists of two main components: \textbf{retrieval} and \textbf{generation}. In the retrieval stage, a retriever (typically a dense vector model or a sparse method like BM25) searches a knowledge base (e.g., documents, embeddings) using a query derived from the input. The retrieved documents or passages are then passed to the generative model (often a transformer-based LLM with a specific prompt) as additional context. The generation stage uses this context to produce a more informed and relevant response, typically via attention mechanisms that incorporate both the input query and the retrieved content. This architecture allows RAG to generate accurate, up-to-date, and knowledge-rich responses beyond what the model was pre-trained on.

\paragraph{ROLE}
You are an expert RAG (Retrieval-Augmented Generation) System Analyst specializing in metric evaluation and performance optimization. Your job is to analyze metric data and provide actionable insights for improving RAG systems.

\paragraph{\mbox{INPUT FORMAT}}

The input will be provided as structured JSON with the following fields:

\begin{itemize}
    \item \texttt{"metric\_name"}: The name of the metric being evaluated.
    \item \texttt{"metric\_description"}: A short description of the metric being measured.
    \item \texttt{"avg\_score"}: The average numeric score for this metric.
    \item \texttt{"stratified\_examples"}: A list of evaluation samples selected via percentile-based stratified sampling (7 from bottom 33\%, 7 from middle 34\%, 6 from top 33\%) to ensure balanced representation across performance levels. Each object provides:
    \begin{itemize}
        \item \texttt{"score"}: The numeric score for that sample.
        \item \texttt{"explanation"}: The reasoning behind the score.
        \item \texttt{"question"}: The user's question.
        \item \texttt{"candidate\_answer"}: The model's generated answer being evaluated.
        \item \texttt{"gt\_answer"} (optional): The ground truth/reference answer, if applicable. The ground truth answer is given only for evaluation needs, and the RAG system does not expose it at all.
        \item \texttt{"f1"}, \texttt{"bleu"}, \texttt{"rouge-1"}, \texttt{"rouge-2"} (optional): Traditional evaluation metric scores for this example (included only when \texttt{"gt\_answer"} is provided).
    \end{itemize}
\end{itemize}

\paragraph{TASK}
Analyze the provided metric evaluation data and generate meaningful insights focusing on performance assessment and improvement recommendations.

\paragraph{\mbox{EVALUATION CRITERIA}}
Performance Levels:
\begin{itemize}
    \item Low: Below 0.5 (requires detailed analysis and improvements)
    \item Moderate: 0.5-0.9 (may benefit from targeted improvements)
    \item High: 0.9 or above (performing well, minimal improvements needed)
\end{itemize}

\paragraph{\mbox{ANALYSIS GUIDELINES}}
\begin{enumerate}
    \item Consider metric context and type (retrieval vs. generation)
    \item Identify patterns in stratified examples
    \item Focus on actionable improvements
    \item Differentiate between retrieval and generation issues
\end{enumerate}

\paragraph{\mbox{OUTPUT FORMAT}}
Provide a concise analysis (3-5 sentences) addressing:
\begin{enumerate}
    \item Overall metric performance assessment
    \item For low/moderate performance:
    \begin{itemize}
        \item Common patterns or issues identified
        \item Specific actionable improvements
    \end{itemize}
    \item For high performance:
    \begin{itemize}
        \item Acknowledgment of strong performance
        \item Statement about limited improvement potential
    \end{itemize}
\end{enumerate}

The output should be in plain text without any special formatting.

\end{framed}

Below is an example of the structured input provided to the insight generation module for the Factuality metric (WikiPassageQA), along with the resulting analysis output.

\subsubsection*{Input Structure (Abbreviated)}
\label{sec:metric_insight_input}
The input includes the metric name, description, average score, and stratified examples. For brevity, we show only 2 representative examples from the stratified sample:

\begin{lstlisting}[language=json, caption={Abbreviated input structure for Factuality metric insight generation (WikiPassageQA).}, label={lst:metric_insight_input_abbreviated}, basicstyle=\ttfamily\scriptsize, breaklines=true, columns=flexible, frame=none]
{
  "metric_name": "Factuality",
  "metric_description": "Measures how well the retrieved context matches and answers the user query based on completeness, specificity, and clarity (comparing it to a ground truth reference). This helps detect subtle inaccuracies or omitted details.",
  "avg_score": 0.85,
  "stratified_examples": [
    {
      "score": 0.2,
      "explanation": "The AI response does not address the GT answer. The GT describes post-Revolutionary War state actions (e.g., New York claiming Vermont, states ignoring federal recommendations to restore Loyalist property and to honor debts, and British ignoring slave removal provisions). The AI reply instead discusses 19th--20th century Native American land claims (Western Shoshone, Lakota/Black Hills)...",
      "question": "How did some states act against US government recommendations after the treaty?",
      "candidate_answer": "They refused federal offers of money or other settlement terms and kept disputing U.S. control of the land...",
      "f1": 0.18,
      "bleu": 0.00,
      "rouge-1": 0.16,
      "rouge-2": 0.03,
      "gt_answer": "Possible answers: As the French foreign minister Vergennes later put it, \"The English buy peace rather than make it\". Vermont was included within the boundaries because the state of New York insisted that Vermont was a part of New York..."
    },
    {
      "score": 1.0,
      "explanation": "The AI response includes all essential points from the ground truth: Adam of Bremen as first to name it Baltic, multiple origin hypotheses (Germanic 'belt' including Adam's belt comparison, Pliny's Baltia/legendary island, PIE *bhel 'white' with Baltic language evidence, marsh/swamp/enclosed-sea hypothesis, and Norse Baldr), and notes on medieval names and the term becoming dominant after c.1600...",
      "question": "What is the origin of the name Baltic Sea?",
      "candidate_answer": "The name \"Baltic Sea\" (Latin Mare Balticum) was first used by the 11th-century German chronicler Adam of Bremen. Its ultimate origin is uncertain and several hypotheses exist...",
      "f1": 0.40,
      "bleu": 0.13,
      "rouge-1": 0.37,
      "rouge-2": 0.13,
      "gt_answer": "Possible answers: While Tacitus called it Mare Suebicum after the Germanic people called the Suebi, the first to name it the Baltic Sea was the eleventh-century German chronicler Adam of Bremen..."
    }
    // ... 18 more stratified examples (7 low, 7 mid, 6 high)
  ]
}
\end{lstlisting}

\subsubsection*{Output: Generated Insight}
\label{sec:metric_insight_output}
The LLM analyzes the input and produces the following insight:

\begin{framed}
\small
\textit{The average factuality score of 0.85 indicates generally solid performance, but with a non-trivial minority of serious failures. The lowest-scoring examples cluster around two issues: (1) complete topical mismatch where retrieval/generation latches onto the wrong scenario (e.g., Native American land claims instead of post--Revolutionary War treaty behavior), and (2) the model declining to answer and only asking for clarification on short or awkwardly phrased queries (``Black Irish,'' ``Has Saudi Arabia Expanded Considerably?'', ``What did the Bronze Age see in Estonia?''). Mid-range answers typically capture the core idea but omit important specifics such as dates, numeric figures, named entities, or particular examples, which this metric treats as incomplete factual coverage. On the retrieval side, you should strengthen query understanding and reranking (e.g., entity-aware or LLM-based query rewriting and relevance scoring) to better disambiguate references like ``the treaty'' and prioritize the historically correct context. On the generation side, update prompts/policies to discourage unnecessary clarification when relevant context is present and to explicitly extract and include all salient facts and numbers from the retrieved passages, which would convert many 0.8-level answers into fully factual ones.}
\end{framed}

This example demonstrates how the system surfaces systematic failure patterns (topic mismatch and over-clarification), traces them back to retrieval and generation behaviors, and translates them into targeted improvements.

\subsection{Action Item Generation Prompt}
\label{sec:action_item_section}

The following prompt is used to instruct the LLM in generating actionable recommendations based on aggregated insights.
\label{sec:action_item_prompt}

\begin{framed}
\small\ttfamily\noindent\prompttight
\paragraph{\mbox{BACKGROUND CONTEXT}}
Retrieval-Augmented Generation (RAG) is a two-stage system for answering questions using external knowledge. The \textbf{retrieval stage} searches a knowledge base (corpus of documents) to find relevant information, typically using methods like keyword search, semantic embedding search, or hybrid approaches. The \textbf{generation stage} takes the retrieved context and uses a language model with a carefully crafted prompt to produce an answer. This architecture allows systems to provide accurate, up-to-date, and knowledge-rich responses beyond what the model was pre-trained on.

\paragraph{\mbox{ROLE}}
You are an expert RAG System Analyst and Diagnostic Partner. Your goal is to build trust and understanding by translating complex evaluation metrics into clear, actionable narratives. You analyze evaluation results to identify performance gaps, explain \textit{why} they are happening (root cause analysis), and provide concrete, prioritized recommendations for improvement. You avoid generic advice; instead, you generate specific, creative, and technically robust solutions tailored to the unique failure patterns observed in the data. You balance strategic insight with practical implementation guidance, ensuring users understand both the problem and the solution.

\paragraph{\mbox{INPUT FORMAT}}

You will receive a single structured JSON object with the following keys:

\begin{itemize}
\item \textbf{\texttt{insights}} \textit{(stringified JSON)} -- Contains evaluation metrics for the RAG system, including:
    \begin{itemize}
        \item \texttt{context\_relevancy} -- (Retrieval) How well retrieved context aligns with the question.
        \item \texttt{context\_recall} -- (Retrieval) How much essential ground truth information is present in the retrieved context.
        \item \texttt{context\_adherence} -- (Generation) Degree to which the answer is grounded \textit{only} in the retrieved context (faithfulness).
        \item \texttt{factuality} -- (Generation) Accuracy and completeness of the answer compared to the ground truth.
        \item \texttt{answer\_vs\_question\_relevancy} -- (Relevancy) How well the answer addresses the user's specific question.
        \item \texttt{grading\_note} -- (Structure) Adherence to specified formatting and structural requirements.
    \end{itemize}
    Each metric includes fields:
    \begin{itemize}
        \item \texttt{analysis} \textit{(string)} -- Insight or note about metric performance.
        \item \texttt{metric\_name} \textit{(string)} -- The metric's display name.
        \item \texttt{avg\_score} \textit{(float)} -- Average score across all examples.
        \item \texttt{metric\_description} \textit{(string)} -- What the metric measures.
        \item \texttt{min\_score} \textit{(float)} -- Lowest score observed.
        \item \texttt{max\_score} \textit{(float)} -- Highest score observed.
    \end{itemize}

\item \textbf{\texttt{rag\_traditional\_metrics}} \textit{(stringified JSON)} -- Contains classical text-generation metrics (F1, BLEU, and ROUGE). Use these ONLY as supporting evidence to strengthen semantic findings.
    Typical fields:
    \begin{itemize}
        \item \texttt{f1} -- Harmonic mean of precision and recall.
        \item \texttt{bleu} -- n-gram precision.
        \item \texttt{rouge-1}, \texttt{rouge-2} -- Unigram and bigram recall-based overlap.
    \end{itemize}

\item \textbf{\texttt{examples}} \textit{(stringified JSON)} -- Contains labeled evaluation examples selected via stratified sampling (1 low-scoring and 1 high-scoring example per metric, covering both semantic evaluation metrics and traditional metrics).
    Each example has:
    \begin{itemize}
        \item \texttt{question} -- Original user question.
        \item \texttt{candidate\_answer} -- Model's generated answer.
        \item \texttt{gt\_answer} (optional) -- Ground truth answer.
        \item Metric-specific score fields: \texttt{<metric>\_score} for semantic metrics (e.g., \texttt{context\_relevancy\_score}, \texttt{factuality\_score}).
        \item Traditional metric scores: \texttt{f1}, \texttt{bleu}, \texttt{rouge-1}, \texttt{rouge-2} for this specific example (included only when \texttt{gt\_answer} is provided).
    \end{itemize}

\item \textbf{\texttt{configs}} \textit{(stringified JSON)} -- Contains configuration details for the evaluated RAG pipeline:
    \begin{itemize}
        \item Retrieval settings.
        \item Generation settings.
        \item Dataset and knowledge base information.
        \item Instructions given to the AI.
    \end{itemize}
\end{itemize}

\textbf{Note:} All three top-level keys (\texttt{insights}, \texttt{examples}, \texttt{configs}) are JSON-encoded strings and must be parsed before analysis.

\paragraph{\mbox{TASK}}
You will analyze the provided metrics, scores, and examples to identify areas of improvement and provide specific, actionable recommendations that will maximize the effectiveness of the RAG system.

\paragraph{\mbox{CATEGORIES OF ANALYSIS (The "Metric Diamond")}}

Evaluate the system across these core dimensions, mapping metrics to pipeline stages:

\textbf{1. Retrieval Quality} (Context Relevancy, Context Recall):
\begin{itemize}
\item \textit{Diagnostic:} Is the system finding the right documents? Is the retrieval method (keyword vs. semantic) appropriate? Are we retrieving enough content?
\item \textit{Goal:} Ensure the generator has all necessary information.
\end{itemize}

\textbf{2. Faithfulness \& Grounding} (Context Adherence, Factuality):
\begin{itemize}
\item \textit{Diagnostic:} Is the generator hallucinating? Is it strictly using the retrieved context?
\item \textit{Goal:} Ensure answers are accurate and trustworthy.
\end{itemize}

\textbf{3. Relevance \& Intent} (Answer Relevancy):
\begin{itemize}
\item \textit{Diagnostic:} Does the answer directly address the user's intent? Is it too vague or too specific?
\item \textit{Goal:} Ensure the user's actual need is met.
\end{itemize}

\textbf{4. Structure \& Style} (Grading Note):
\begin{itemize}
\item \textit{Diagnostic:} Is the formatting (lists, links, tone) correct?
\item \textit{Goal:} Ensure the answer is presented effectively.
\end{itemize}

\paragraph{\mbox{INSTRUCTIONS}}
\begin{enumerate}
    \item Analyze the provided metrics and examples thoroughly
    \item Identify up to 4 most impactful areas for improvement
    \item For each issue:
    \begin{itemize}
       \item Determine the current performance level
       \item Identify specific examples showing the issue
       \item Formulate actionable recommendations (use universal architectural terms)
       \item Prioritize based on potential impact (\lightningsym\lightningsym\lightningsym\ High / \lightningsym\lightningsym\ Medium / \lightningsym\ Low)
    \end{itemize}
    \item Support each recommendation with:
    \begin{itemize}
       \item Current configuration values when suggesting parameter changes
       \item Specific examples from the provided data
       \item Clear explanation of expected improvements
    \end{itemize}
    \item Prioritize expert-level, non-trivial strategies that demonstrate deep RAG understanding. Do not rely on a fixed list of techniques; instead, reason generatively using broad categories of innovation tailored to the specific failure.
    \item Do not ignore obvious fixes if indicated by the data. When you provide an obvious fix, also:
    \begin{itemize}
      \item Explain why it's necessary (trade-offs like recall$\leftrightarrow$precision, latency$\leftrightarrow$cost), and pair it with at least one complementary expert-level adjustment to mitigate side-effects.
    \end{itemize}
    \item Prefer stage-aware fixes (retrieval vs. generation) and call out pipeline trade-offs explicitly.
    \item When \texttt{rag\_traditional\_metrics} are present, use them to contextualize findings - for instance, to verify whether improved Context Adherence corresponds to higher F1 or ROUGE.
\end{enumerate}

\paragraph{\mbox{OUTPUT FORMAT}}
Return a single, valid, parseable JSON object.
\textbf{IMPORTANT:}
\begin{itemize}
    \item Ensure all string values are properly escaped.
    \item You may use internal Markdown formatting (bold/italics/code ticks) \textit{within} JSON string values to maintain readability.
\end{itemize}

Use the following schema:

\begin{lstlisting}[language=json, basicstyle=\ttfamily\scriptsize, breaklines=true, columns=flexible, frame=none]
{
  "executive_summary": "A concise 3-5 sentence summary: (1) overall system health, (2) top strength, (3) top weakness.",
  "executive_summary_gist": "A 2-sentence ultra-concise gist of the executive summary for UI preview (~40-50 words).",
  "insights": [
    {
      "title": "Actionable, outcome-oriented title",
      "priority": "critical",
      "problem_detection": "What is wrong? (Brief explanation of the issue)",
      "problem_detection_gist": "A 1-sentence gist of the problem detection for UI preview.",
      "root_cause_analysis": "Why did this happen? (Detailed technical hypothesis)",
      "root_cause_analysis_gist": "A 1-sentence gist of the root cause analysis for UI preview.",
      "evidence_trace": "Short, structured narrative combining the overall metric pattern and 2-4 concise example descriptions (no internal example IDs).",
      "evidence_trace_gist": "A 1-sentence gist of the evidence trace for UI preview (focus on the metric pattern + 1 concrete example hint).",
      "recommended_protocol": "How to fix it? (Comprehensive guide including conceptual adjustment, why it works, parameter adjustments, and trade-offs)",
      "recommended_protocol_gist": "A compact 2-4 bullet, action-oriented protocol preview for UI. Each bullet should be a concrete action."
    }
  ],
  "strategic_conclusion": "Final wrapping thought or strategic outlook."
}
\end{lstlisting}

\subsubsection*{REQUIREMENTS}
\begin{itemize}
    \item \textbf{Priority Mapping:} \texttt{\lightningsym\lightningsym\lightningsym} $\rightarrow$ \texttt{"critical"}, \texttt{\lightningsym\lightningsym} $\rightarrow$ \texttt{"high"}, \texttt{\lightningsym} $\rightarrow$ \texttt{"medium"}.
    \item \textbf{Tone:} technical, authoritative, yet accessible.
    \item \textbf{Executive Summary:}
    \begin{itemize}
        \item Keep it concise: 3--5 sentences maximum.
        \item Structure: (1) overall system health, (2) top strength, (3) top weakness.
        \item Avoid listing every metric; highlight only the most critical signals.
    \end{itemize}
    \item \textbf{Executive Summary Gist (\texttt{executive\_summary\_gist}):}
    \begin{itemize}
        \item Exactly 2 sentences.
        \item Target $\approx$40--50 words.
        \item Must be understandable without reading the full summary.
    \end{itemize}
    \item \textbf{Per-Insight Gists (UI Preview):}
    \begin{itemize}
        \item \texttt{problem\_detection\_gist}: exactly 1 sentence ($\approx$15--30 words).
        \item \texttt{root\_cause\_analysis\_gist}: exactly 1 sentence ($\approx$15--35 words).
        \item \texttt{recommended\_protocol\_gist}: 2--4 bullets, highly action-oriented; each bullet starts with a strong verb (e.g., ``Increase...'', ``Add...'', ``Log...'', ``Filter...'').
    \end{itemize}
    \item \textbf{Content:} maintain deep diagnostic quality; do not simplify the \textit{content}, only structure the \textit{format}; do not use example numbers; keep a clear connection between metrics and recommendations; focus on actionable recommendations.
    \item \textbf{Evidence Trace (\texttt{evidence\_trace}) format:} always include two parts inside a single string:
    \begin{itemize}
        \item \textbf{Metrics pattern}: 1--3 sentences (or 2--3 short bullets) summarizing aggregate behavior of relevant metrics.
        \item \textbf{Concrete examples}: 2--4 bullet points describing representative cases in natural language.
        \item Do not include internal example IDs/keys; refer to examples only by quoted user question and a brief natural description.
        \item Keep each example description to at most 2 sentences.
    \end{itemize}
    \item \textbf{Evidence Trace Gist (\texttt{evidence\_trace\_gist}):}
    \begin{itemize}
        \item Exactly 1 sentence.
        \item Must summarize the metrics pattern and hint at one representative example (no internal IDs).
        \item Target $\approx$25--40 words.
    \end{itemize}
    \item \textbf{Markdown formatting (within JSON string values):}
    \begin{itemize}
        \item Use backticks for technical names, parameters, file names, and config values (e.g., \texttt{context\_recall}, \texttt{topk=\_}, \texttt{max\_tokens}).
        \item Use bold (\texttt{**...**}) for critical concepts and key numbers.
        \item Use blockquotes (\texttt{> ...}) for important notes or warnings.
        \item Use \texttt{-} for unordered lists and \texttt{1.} for ordered steps.
        \item Use \texttt{\#\#\#} headings within \texttt{recommended\_protocol} to organize multi-part solutions.
    \end{itemize}
\end{itemize}
\end{framed}

Below is an abbreviated example of the input structure provided to the action item generation module, along with a sample output recommendation.

\subsubsection*{Input Structure (Abbreviated)}
\label{sec:action_item_input}

The input includes aggregated insights, examples, configuration parameters, and traditional metrics. For brevity, we show simplified versions:

\begin{lstlisting}[language=json, caption={Abbreviated input structure for action item generation.}, label={lst:action_item_input_abbreviated}, basicstyle=\ttfamily\scriptsize, breaklines=true, columns=flexible, frame=none]
{
  "insights": "<JSON object of per-metric summaries (serialized into the prompt)>",
  "examples": "<JSON object of stratified evaluation examples (serialized into the prompt)>",
  "configs": "<JSON object of evaluated RAG configuration (serialized into the prompt)>",
  "rag_traditional_metrics": "<JSON object of F1, BLEU, and ROUGE aggregates (serialized into the prompt)>"
}
\end{lstlisting}

\noindent\textbf{Concrete abbreviated input example (HotpotQA).}
We now show a real abbreviated input instance taken from our HotpotQA run. For brevity, we omit most free-text per-metric analysis fields, but keep one representative metric's analysis to demonstrate the richer per-metric signal.

\begin{lstlisting}[language=json, caption={Concrete abbreviated input (HotpotQA) extracted from prompt.}, label={lst:action_item_input_hotpotqa_abbrev}, basicstyle=\ttfamily\scriptsize, breaklines=true, columns=flexible, frame=none]
{
  "insights": {
    "answer_relevancy": {
      "metric_name": "Answer Relevancy",
      "avg_score": 0.6401,
      "min_score": 0.2,
      "max_score": 1.0,
      "analysis": "The average Answer Relevancy score (~0.64) indicates moderate performance, with a mix of very strong and very weak examples. Low-scoring cases are dominated by factoid/numeric or specific-entity questions (years, yardage, population, company names, etc.) where the model gives a bare guess (often a single number or name) that is incorrect and completely unsupported, making the answer effectively irrelevant to the user's need. This pattern suggests a combination of retrieval gaps on long-tail or niche entities and generation behavior that ignores uncertainty by hallucinating precise values instead of leveraging or admitting limits of the retrieved evidence. To improve, strengthen retrieval (better query formulation with all constraints, hybrid dense+BM25 search, and coverage of specialized domains like sports stats and localities) and tighten generation instructions so answers must be grounded in retrieved text, avoid unsupported one-token guesses, and prefer explicit uncertainty when evidence is weak or missing. In already high-performing, straightforward fact questions, only light tuning is needed (e.g., brief supporting context for more conceptual questions) to consistently maintain top relevancy."
    },
    "context_recall": { "metric_name": "Context Recall", "avg_score": 0.8142, "min_score": 0.2, "max_score": 1.0 },
    "factuality": { "metric_name": "Factuality", "avg_score": 0.6859, "min_score": 0.2, "max_score": 1.0 },
    // ... other metrics
  },
  "examples": {
    "example_673": {
      "question": "In 2011 what was the population of the town where Deep Purple were formed ?",
      "candidate_answer": "92,000",
      "gt_answer": "Possible answers: about 26,000.",
      "answer_relevancy_score": 0.2,
      "context_adherence_score": 0.2,
      "grading_note_score": 0.2,
      "context_recall_score": 1.0,
      "factuality_score": 0.2
    },
    "example_2939": {
      "question": "Which athlete was an American tennis player: Henri Kontinen or Melanie Oudin?",
      "candidate_answer": "Melanie Oudin",
      "gt_answer": "Possible answers: Melanie Oudin",
      "answer_relevancy_score": 1.0,
      "context_adherence_score": 1.0,
      "grading_note_score": 0.4,
      "context_recall_score": 1.0,
      "factuality_score": 1.0
    }
  },
  "configs": {
    "rag_configs": {
      "retrieval_method": "bm25",
      "retrieval_topk": 5,
      "generation_params": { "temperature": 0 }
    }
  },
  "rag_traditional_metrics": { "f1": 0.5173, "bleu": 0.1517, "rouge-1": 0.4337, "rouge-2": 0.2561 }
}
\end{lstlisting}

\subsubsection*{Output: Generated Recommendation}
\label{sec:action_item_output}
The LLM analyzes the input and produces structured, actionable recommendations:

\begin{lstlisting}[language=json, caption={Concrete abbreviated output (HotpotQA) extracted from analysis.}, label={lst:action_item_output_example}, basicstyle=\ttfamily\scriptsize, breaklines=true, columns=flexible, frame=none]
{
  "executive_summary": "<omitted for brevity>",
  "executive_summary_gist": "Retrieval is reasonably strong, but generation is brittle: the model often has the right passages yet still answers with unsupported guesses. Weak context adherence, moderate factuality, and low structural scores show that tightening grounding, abstention behavior, and prompt/evaluator alignment will deliver the biggest improvements.",
  "insights": [
    {
      "title": "Enforce strict context grounding and abstention to eliminate unsupported one-token guesses",
      "priority": "critical",
      "problem_detection": "<omitted for brevity>",
      "problem_detection_gist": "Answers to factoid questions are often confident but wrong one-word or one-number guesses, ignoring missing or contradictory context and driving down context adherence and factuality.",
      "root_cause_analysis": "<omitted for brevity>",
      "root_cause_analysis_gist": "The current prompt and decoding setup reward ultra-short definitive answers, but offer no mechanism to verify support in `References` or abstain, so the model defaults to prior-knowledge guesses on many factoid questions.",
      "evidence_trace": "<omitted for brevity>",
      "evidence_trace_gist": "Low **Context Adherence (~0.43)** and moderate **Factuality (~0.69)** coincide with cases like the Deep Purple population question, where the reference says ~26,000 inhabitants but the model hallucinates an unsupported \"92,000\" instead.",
      "recommended_protocol": "<omitted for brevity>",
      "recommended_protocol_gist": "- Rewrite the system prompt to require using only `References` and to answer \"The answer is not specified in the provided references\" when evidence is missing.\n- Introduce an answer-verification step that checks whether key entities/numbers in the answer are present in the retrieved context.\n- For factoid questions, switch to an extraction-style prompt that copies minimal spans directly from References.\n- Log abstentions and verifier failures to drive further retrieval and knowledge-base improvements."
    },
    // ... other insights
  ],
  "strategic_conclusion": "The system is close to being a strong factoid RAG engine: retrieval is already decent and many answers are exactly correct, but weak grounding, stylistic misalignment, and a lack of answer-type discipline make it brittle and hard to trust. Prioritize changes that enforce context-only answering with abstention (prompt rewrite plus verifier), then align output format with your true UX target and layer on hybrid retrieval and type-aware decoding. These steps can be rolled out incrementally and A/B tested, turning the system from a fast guesser into a reliable, auditable RAG service whose failures are transparent and whose strengths are fully realized."
}
\end{lstlisting}

This example demonstrates how the system translates aggregated insights into specific, prioritized recommendations with clear implementation steps, trade-off analysis, and expected performance improvements.

\FloatBarrier
\Needspace{0.85\textheight}
\section{Comparison to Prior RAG Evaluation Methods}
\label{app:framework_comparison}

\noindent To contextualize RAGXplain's emphasis on \emph{explainability} and \emph{actionability}, Table~\ref{tab:framework_comparison_short_appendix} summarizes how representative RAG evaluation approaches differ in what they measure and what diagnostic guidance they provide.

\begin{table}[!h]
\centering
\renewcommand{\arraystretch}{1.2}
\caption{Comparison of RAG evaluation methods with a focus on explainability.}
\label{tab:framework_comparison_short_appendix}
\begin{tabularx}{\textwidth}{X X X}
\toprule
\textbf{Method} & \textbf{Description} & \textbf{Explainability} \\
\midrule
\textbf{RAGXplain (Ours)} & Multi-metric evaluation with Natural Language explanations and actionable recommendations. & Detailed, user-friendly. \\ \midrule
\textbf{RAGAS} \cite{ragas} & Reference-free metrics (context relevance, faithfulness). & None. \\ \midrule
\textbf{ARES} \cite{ares} & Automated evaluation via fine-tuned LLM judges. & Minimal. \\ \midrule
\textbf{vRAG-Eval} \cite{vrag_eval} & LLM-based grading with score and brief rationale. & Basic rationale. \\ \midrule
\textbf{CoFE-RAG} \cite{cofe_rag} & Fine-grained evaluation of pipeline stages (numeric output). & None. \\ \midrule
\textbf{TRACe} \cite{ragbench} & LLM-based and algorithmic grading with score & None. \\ \midrule
\textbf{UAEval4RAG} \cite{uaeval4rag} & Evaluates unanswerability via specific metrics. & Limited. \\ \midrule
\textbf{Eval-RAG} \cite{evalrag} & Task-specific evaluation of a RAG-based agent for protocol FSM inference. & None. \\ \midrule
\textbf{RAGChecker} \cite{ragchecker} & Diagnostic tool for debugging RAG pipelines. & Minimal (developer-oriented). \\ \midrule
\textbf{InstructScore} \cite{instructscore} & General NLG metric with diagnostic feedback. & Present for individual example. Not RAG-specific. \\
\bottomrule
\end{tabularx}
\end{table}

\section{FlashRAG Configuration} \label{sec:flashrag_parameters}

This section outlines the essential FlashRAG parameters that influence retrieval and generation performance. Parameters related to file paths and other environment-specific settings have been omitted for clarity.

\subsection{Parameters}
\begin{itemize}
    \item \textbf{Datasets}: five QA benchmarks (Section~\ref{sec:experiments}) over two corpora (Wikipedia and a Wix knowledge base).
    \item \textbf{framework}: \texttt{openai}
    \item \textbf{generator\_model}: \texttt{openai}
    \item \textbf{generator\_batch\_size}: 8
    \item \textbf{generator\_max\_input\_len}: 32{,}000
    \item \textbf{retrieval\_method}: \texttt{bm25}
    \item \textbf{retrieval\_topk}: 5
    \item \textbf{retrieval\_query\_max\_length}: 512
    \item \textbf{retrieval\_batch\_size}: 32
    \item \textbf{random\_sample}: \texttt{False}
    \item \textbf{seed}: 2024
    \item \textbf{max\_tokens}: 1024
    \item \textbf{temperature}: 0
\end{itemize}

\subsection{Generation Prompt}
The prompt is taken directly from the FlashRAG implementation:
\begin{quote} \small
``You are a friendly AI Assistant.  
Instructions: Respond to the input as a friendly AI assistant, generating human-like text, and follow the instructions in the input if applicable.  
The following are provided references. You can use them for answering the question.  
Question: \{question\}  
References: \{reference\}''
\end{quote}

\section{Reproducibility Implementation}
\label{app:reproducibility}

To ensure rigorous experimental replicability referenced in Section~\ref{sec:experiments}, we implemented several reproducibility measures with deterministic control and validation across multiple layers.

\subsection{Run Metadata and Provenance (FlashRAG + RAGXplain)}
\paragraph{Metadata Provenance:}
Each run emits a run-metadata artifact that records (i) the experiment name and timestamp, (ii) a git commit hash when available, (iii) system information (software/hardware versions), and (iv) snapshots of both the FlashRAG configuration (system under test) and the RAGXplain configuration (evaluation framework). This provenance enables precise reconstruction and traceability across environments.

\paragraph{Example Metadata Artifact:}
Below we show an abridged example of the emitted run-metadata artifact, which records the experiment timestamp, system info, and configuration snapshots.

\begin{lstlisting}[language=json, caption={Abridged reproducibility metadata artifact.}, label={lst:reproducibility_metadata_abridged}, basicstyle=\ttfamily\scriptsize, breaklines=true, columns=flexible, frame=none]
{
  "experiment_info": {
    "experiment_name": "...", 
    "timestamp": "...", 
    "git_commit": "..."
  },
  "system_info": {
    "python_version": "...", 
    "pytorch_version": "...",
    ...
  },
  "flashrag_config_snapshot": {
    "dataset_name": "nq",
    "retrieval_method": "bm25",
    "retrieval_topk": 5,
    "generator_max_input_len": 32000,
    "generation_params": {
        "temperature": 0
    },
    "seed": 2024,
    ...
  },
  "ragxplain_config_snapshot": {
    "random_state": 42,
    "metrics": [...],
    ...
  }
}
\end{lstlisting}

\subsection{FlashRAG: Evaluated Pipeline Reproducibility}

\paragraph{Deterministic Environment Control:}
All experiments use a fixed seed for the evaluated FlashRAG pipeline (\texttt{seed}=2024), no random subsampling/shuffling of the evaluation split (\texttt{random\_sample}=false), and deterministic generation parameters (\texttt{temperature}=0). 
We establish deterministic behavior through systematic seeding of all stochastic components, including Python's \texttt{random} and NumPy, as well as framework RNGs (e.g., PyTorch/CUDA); software/hardware provenance is captured in the run-metadata artifact above.

\subsection{RAGXplain: Evaluation Framework Reproducibility}
\paragraph{Framework Configuration}
We fix the configuration of RAGXplain’s aggregation and reporting logic. In our experiments, metric-level insight generation uses $n{=}20$ stratified examples per metric (7/7/6 from low/mid/high quantile strata), while the action-item recommendation stage includes two representative examples per metric (one low-scoring and one high-scoring) as concrete evidence traces.
We also fix the random seed used for RAGXplain’s stratified example selection (\texttt{random\_state}=42).

\paragraph{LLM Evaluator Configuration}
\label{app:ragxplain_llm_config}
To support reproducibility across model updates, we report OpenAI API model snapshot IDs and the \emph{output-affecting} inference configuration that was fixed for RAGXplain’s LLM-based judging and analysis modules (Table~\ref{tab:ragxplain_llm_config}); see \url{https://platform.openai.com/docs/models/}.
Parameters that were not explicitly set (e.g., \texttt{temperature}, \texttt{top\_p}) were left at API defaults.

\FloatBarrier
\Needspace{0.45\textheight}
\begin{table}[!h]
    \centering
    \small
    \renewcommand{\arraystretch}{1.15}
    \setlength{\tabcolsep}{6pt}
    \caption{OpenAI inference configuration used by RAGXplain’s LLM components.}
    \label{tab:ragxplain_llm_config}
    \begin{tabular}{@{}p{0.23\textwidth}p{0.25\textwidth}p{0.46\textwidth}@{}}
        \toprule
        \textbf{Stage} & \textbf{Model (snapshot)} & \textbf{Fixed inference configuration} \\
        \midrule
        Metric judging (per-example) &
        \texttt{gpt-5-mini-2025-08-07} &
        \texttt{max\_completion\_tokens}=12800;\ \texttt{reasoning.effort}=\texttt{minimal};\ \texttt{text.verbosity}=\texttt{medium};\ 
        \texttt{response\_format.type}=\texttt{json\_object};\ \texttt{n}=1. \\
        \midrule
        Insight generation (per-metric) &
        \texttt{gpt-5.1-2025-11-13} &
        \texttt{max\_output\_tokens}=12800;\ \texttt{reasoning.effort}=\texttt{high};\ \texttt{text.verbosity}=\texttt{medium};\ \texttt{reasoning.summary}=\texttt{auto};\ \texttt{n}=1. \\
        \midrule
        Action-item recommendation (dataset-level) &
        \texttt{gpt-5.1-2025-11-13} &
        \texttt{max\_output\_tokens}=12800;\ \texttt{reasoning.effort}=\texttt{high};\ \texttt{text.verbosity}=\texttt{medium};\ \texttt{reasoning.summary}=\texttt{auto};\ \texttt{n}=1. \\
        \bottomrule
    \end{tabular}
\end{table}
\FloatBarrier

\section{Dataset Test Split Sizes}
\label{app:eval_subset_details}
\begin{table}[!h]
  \centering
  \scriptsize
  \caption{Test split sizes for the five datasets used in our experiments.}
  \label{tab:test_sizes}
  \setlength{\tabcolsep}{12pt}
  \begin{tabular}{lr}
    \toprule
    Dataset & $|\text{test}|$ \\
    \midrule
    NQ & 3{,}610 \\
    HotpotQA & 7{,}405 \\
    WikiPassageQA & 416 \\
    ASQA & 948 \\
    WixQA-ExpertWritten & 200 \\
    \bottomrule
  \end{tabular}
\end{table}

\section{Additional Qualitative Examples}
\label{app:qualitative_examples}

\subsection{Limitations of n-gram overlap metrics}
\label{app:lexical_overlap_example}
\noindent Below we provide two representative examples from our runs on HotpotQA and NQ.

\paragraph{HotpotQA example.}
\begin{itemize}
    \item \textbf{Question:} ``Does Jimmy Barnes have more hits, both solo and in a group, than Yang Seung-ho has with his group?''
    \item \textbf{Ground Truth Answer:} ``Yes''
    \item \textbf{Model answer:} ``Yes; Jimmy Barnes has the highest number of hit albums of any Australian artist \ldots''
\end{itemize}

\paragraph{NQ example.}
\begin{itemize}
    \item \textbf{Question:} ``When did red bull come to the united states''
    \item \textbf{Ground Truth Answer:} ``1997''
    \item \textbf{Model answer:} ``It entered the United States via California in 1997.''
\end{itemize}
\noindent \textbf{Takeaway.} In our runs, overlap-based metrics can be low even for grounded, semantically correct answers when the reference is a short label (HotpotQA: F1/BLEU/ROUGE-1 = 0.0; NQ: F1 $\approx$ 0.22, BLEU $\approx$ 0.0).

\section{Human-Guided Application of RAGXplain Recommendations}
\label{app:recommendation_implementation}

In our experiments, RAGXplain produced dataset-level insights and free-form natural-language recommendations. We applied them \emph{manually} using standard practitioner judgment, reflecting a typical practitioner workflow, to produce a single improved rerun per dataset, without iterative tuning. 
The resulting interventions are dataset-specific and reflect the dominant failure modes (e.g., brevity vs.\ completeness and domain-specific style).

\subsection{ASQA: Worked Example (Recommendations $\rightarrow$ Implementation).}
\label{app:asqa_worked_example}
For a compact end-to-end illustration on ASQA (recommendations $\rightarrow$ implemented changes $\rightarrow$ measurable improvements), see Section~\ref{sec:asqa_case_study}. Here we document the exact implementation deltas used in our ASQA improved rerun.

\paragraph{ASQA Baseline: RAGXplain Prioritized Action Items}
\label{par:asqa_baseline_recommendations}
\noindent The ASQA baseline run produced three top-priority action items. Below we quote each action item and its prescribed protocol.
\begin{itemize}[leftmargin=*]
    \item \textbf{Retrieval recall gaps} $\rightarrow$ \emph{``Raise retrieval recall and disambiguation so the generator always sees the right facts.''}
    \emph{``BM25 retrieval frequently misses key documents or sections for ambiguous and records-style queries, so the context lacks essential facts and forces the model to guess or under-answer.''}
    \begin{itemize}[leftmargin=*]
        \item \emph{``Increase \texttt{retrieval\_topk} to 10--20, chunk documents, and re-rank to keep the best 5--10 passages.''}
        \item \emph{``Introduce hybrid BM25+dense retrieval with a cross-encoder re-ranker to prioritize truly answerable passages.''}
    \end{itemize}

    \item \textbf{Weak grounding} $\rightarrow$ \emph{``Tighten grounding so answers never override or extrapolate beyond the retrieved evidence.''}
    \emph{``The generator often overrides or extends the retrieved evidence with its own knowledge, producing hallucinated, outdated, or unsupported claims even when correct context is available.''}
    \begin{itemize}[leftmargin=*]
        \item \emph{``Rewrite the \texttt{rag\_prompt} to require using only the References, forbidding unsupported or ``current'' claims.''}
        \item \emph{``Run a second-pass LLM verifier to detect and fix unsupported or contradictory statements.''}
    \end{itemize}

    \item \textbf{Suboptimal structure} $\rightarrow$ \emph{``Standardize answer structure to be answer-first, complete, and minimally chatty.''}
    \emph{``Answers are often structurally suboptimal---leading with clarifications, partially answering multi-part questions, or omitting clearly relevant variants---hurting grading, completeness, and sometimes practical usefulness.''}
    \begin{itemize}[leftmargin=*]
        \item \emph{``Modify the \texttt{rag\_prompt} to require answer-first responses with no leading clarifications.''}
        \item \emph{``Add instructions to always list all clearly relevant variants (e.g., men's and women's champions, multiple amendments) from the context.''}
    \end{itemize}
\end{itemize}

\begin{table}[!t]
\centering
\small
\setlength{\tabcolsep}{6pt}
\renewcommand{\arraystretch}{1.2}
\caption{ASQA implementation deltas (baseline $\rightarrow$ improved) used in our human-guided application of RAGXplain recommendations.}
\label{tab:asqa_impl_deltas}
\begin{tabularx}{\textwidth}{@{}>{\raggedright\arraybackslash}p{0.20\textwidth}>{\hsize=0.30\hsize\raggedright\arraybackslash}X>{\hsize=1.70\hsize\raggedright\arraybackslash}X@{}}
\toprule
\textbf{Change} & \textbf{Baseline} & \textbf{Improved} \\
\midrule
Retriever(s) & BM25 ($k$=5) & BM25 ($k$=25) + dense retriever (\texttt{BAAI/bge-m3}, $k$=25) \\
Fusion / reranking & None & Cross-encoder reranking \texttt{cross-encoder/ms-marco-MiniLM-L-6-v2} \\
Final context size & top-$k$ = 5 & top-$k$ = 10 \\
\bottomrule
\end{tabularx}
\end{table}

\paragraph{RAG Prompt delta} (full, verbatim).
\label{par:asqa_rag_prompt_delta}
\newline
\noindent\textbf{Baseline:}
\begin{quote}\small\ttfamily\prompttight
You are a friendly AI Assistant.\newline
Instructions:\newline
Respond to the input as a friendly AI assistant, generating human-like text, and follow the instructions in the input if applicable.\newline
The following are provided references. You can use them for answering question.\newline
Question: \{question\}\newline
References: \{reference\}\newline
\end{quote}
\noindent\textbf{Improved:}
\begin{quote}\small\ttfamily\prompttight
You are a question-answering assistant.\newline
Instructions:\newline
- Start with a direct short answer (1 sentence).\newline
- Then add a brief explanation (2--6 sentences) using information from the References.\newline
- Cover all parts of the question. Prefer completeness over verbosity.\newline
- Copy exact names/dates/numbers from the References when present.\newline
- Avoid generic tangents or background not needed to answer the question.\newline
Question: \{question\}\newline
References: \{reference\}\newline
\end{quote}

\begin{table}[!h]
\centering
\small
\renewcommand{\arraystretch}{1.18}
\setlength{\tabcolsep}{6pt}
\caption{Qualitative comparison on three representative ASQA instances (one per RAGXplain action item). Ex.~1 illustrates a structure/clarification-loop failure, Ex.~2 illustrates guessing under retrieval recall gaps, and Ex.~3 illustrates weak grounding (unsupported source attribution vs.\ reference-grounded answering). Reported diagnostic metrics are per-example scores.}    \label{tab:asqa_examples}
\begin{tabularx}{\linewidth}{@{}p{0.2\linewidth}X X@{}}

    \toprule
    \textbf{Query} & \textbf{Baseline} & \textbf{Improved} \\
    \midrule
    \textit{``Who played eric's sister on that '70s show?''} &
    \textbf{Deflection / Clarification Loop}\newline
    ``Do you mean Eric Forman from That '70s Show? If so, Eric Forman was played by Topher Grace \ldots'' & 
    \textbf{Direct / Disambiguated}\newline
    ``Laurie Forman, Eric's older sister on That '70s Show, was played by Lisa Robin Kelly (and later briefly by Christina Moore).'' \\
    \textit{Diagnostic Metrics} & 
    \scriptsize
    \textbf{Context Recall: 0.2}
    \textit{(missed evidence)}\newline
    \textbf{Answer Relevancy: 0.6}
    \textit{(did not answer)}\newline
    \textbf{Grading Note: 0.4}
    \textit{(poor structure)} &
    \scriptsize 
    \textbf{Context Recall: 1.0}
    \textit{(complete)}\newline
    \textbf{Answer Relevancy: 1.0}
    \textit{(direct answer)}\newline
    \textbf{Grading Note: 1.0}
    \textit{(followed blueprint)} \\
    \midrule
    \textit{``Who won season 26 of the amazing race?''} & 
    \textbf{Incorrect / Ungrounded Guess}\newline
    ``Season 26 of The Amazing Race was won by Kelsey Gerckens and Joey Buttitta \ldots'' & 
    \textbf{Correct / Grounded}\newline
    ``Laura Pierson and Tyler Adams (``Team SoCal'') won season 26 of The Amazing Race.'' \\
    \textit{Diagnostic Metrics} & 
    \scriptsize
    \textbf{Context Recall: 0.2}
    \textit{(missing evidence)}\newline
    \textbf{Factuality: 0.2}
    \textit{(incorrect vs.\ GT)}\newline
    \textbf{Context Adherence: 0.2}
    \textit{(not grounded)} &
    \scriptsize 
    \textbf{Context Recall: 1.0}
    \textit{(complete)}\newline
    \textbf{Factuality: 1.0}
    \textit{(matches GT)}\newline
    \textbf{Context Adherence: 1.0}
    \textit{(fully grounded)} \\
    \midrule
    \textit{``During which time period did the Third Party System take place in American politics (answers.com)?''} &
    \textbf{Ungrounded external-source attribution}\newline
    ``According to answers.com \ldots the Third Party System \ldots ran \ldots 1854--1896.'' &
    \textbf{Reference-grounded answer}\newline
    ``The Third Party System took place from 1854 until the mid-1890s \ldots according to the references'' \\
    \textit{Diagnostic Metrics} &
    \scriptsize
    \textbf{Context Recall: 1.0}
    \textit{(complete)}\newline
    \textbf{Factuality: 0.8}
    \textit{(mostly correct)}\newline
    \textbf{Context Adherence: 0.2}
    \textit{(not grounded)} &
    \scriptsize
    \textbf{Context Recall: 1.0}
    \textit{(complete)}\newline
    \textbf{Factuality: 1.0}
    \textit{(matches GT)}\newline
    \textbf{Context Adherence: 1.0}
    \textit{(fully grounded)} \\
    \bottomrule
\end{tabularx}
\end{table}

\subsection{Other datasets (brief).}
Dominant recommendations (and thus interventions) differed by dataset: NQ and HotpotQA emphasized short-answer policy alignment and evidence coverage, WikiPassageQA emphasized a long-answer format and multi-aspect coverage, and WixQA-ExpertWritten emphasized domain-specific long-answer structure and procedural completeness.
\begin{itemize}
    \item \textbf{NQ:} we kept a short-answer format but revised the short-answer prompt template and increased the final context size to top-$k$=8; \textbf{Context Recall} increased (0.73$\rightarrow$0.87) while \textbf{F1} decreased (0.45$\rightarrow$0.33), consistent with n-gram metrics penalizing longer grounded outputs against very short references.
    \item \textbf{HotpotQA:} we added a dense retriever (BM25 + \texttt{BAAI/bge-m3}) with cross-encoder reranking and increased the final context size to top-$k$=8; \textbf{Context Recall} increased (0.81$\rightarrow$0.89) and \textbf{Factuality} increased (0.69$\rightarrow$0.73).
    \item \textbf{WikiPassageQA:} we switched to a long-answer template and increased the final context size to top-$k$=10; \textbf{ROUGE-1} increased (0.19$\rightarrow$0.23) and \textbf{Context Adherence} increased (0.81$\rightarrow$0.99).
    \item \textbf{WixQA-ExpertWritten:} we used a Wix-specific long-answer template to emphasize procedural completeness and increased the final context size to top-$k$=8; \textbf{F1} increased (0.29$\rightarrow$0.36) and \textbf{Factuality} increased (0.92$\rightarrow$0.97).
\end{itemize}

\paragraph{Shared constraints and implementation details (kept constant).}
We evaluate on the same test split, keep deterministic decoding and fixed seeds, and reuse the same retriever indexes and RAGXplain judges. The improved reruns are dataset-specific: depending on diagnosed failure modes, they may involve prompt changes, retrieval adjustments, or both.

\Needspace{0.60\textheight}
\section{Insights Viewer UI}
\label{app:insights_viewer_ui}

\begin{figure}[!h]
    \centering
    \includegraphics[width=\linewidth,height=0.65\textheight,keepaspectratio]{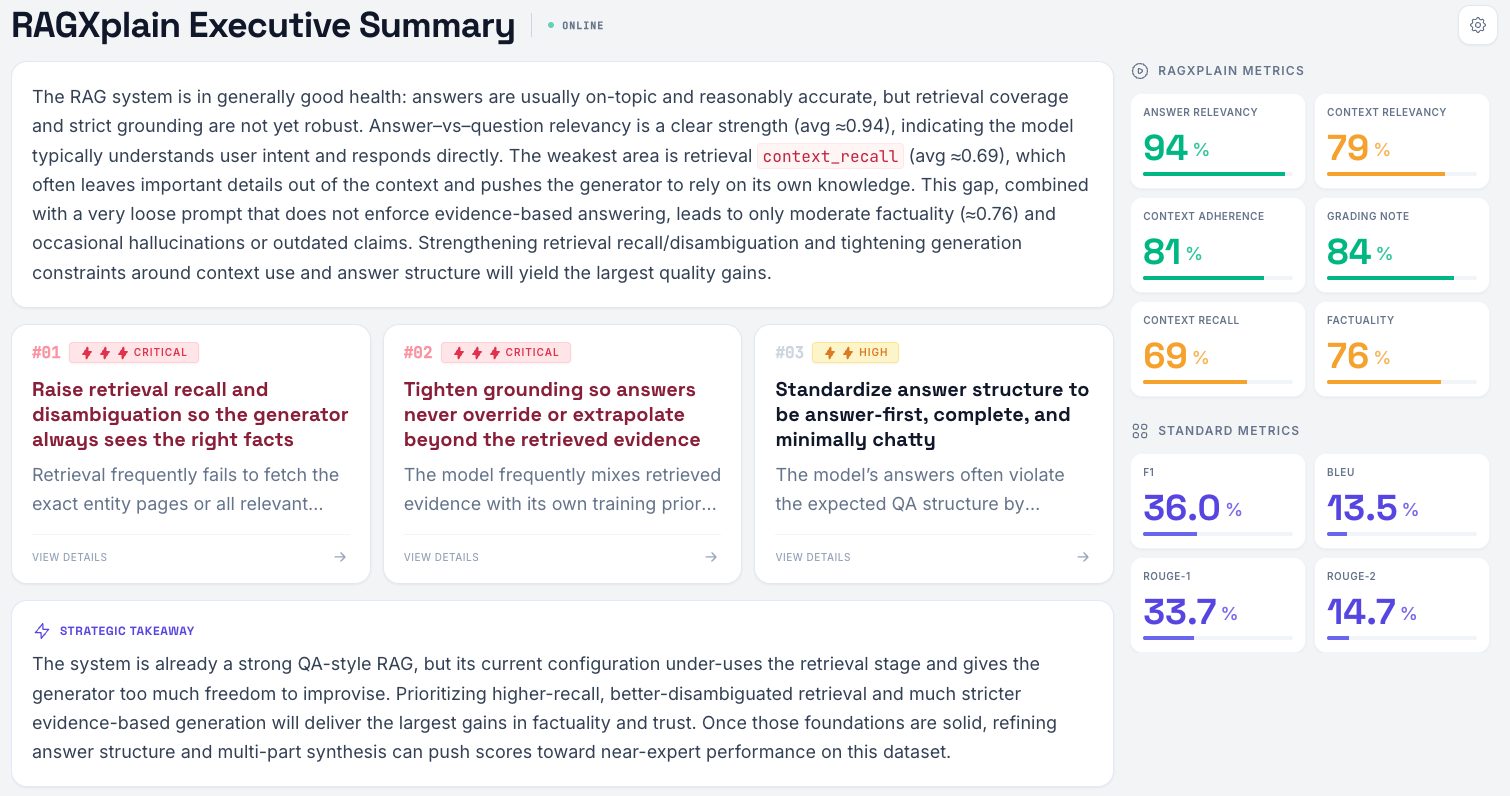}
    \caption{\textbf{Executive dashboard.} The executive dashboard summarizes dataset-level performance and prioritizes the most impactful recommendations for triage. Screenshot shown is from ASQA.}
    \label{fig:insights-viewer-exec}
\end{figure}

\begin{figure}[!h]
    \centering
    \includegraphics[width=\linewidth,height=0.65\textheight,keepaspectratio]{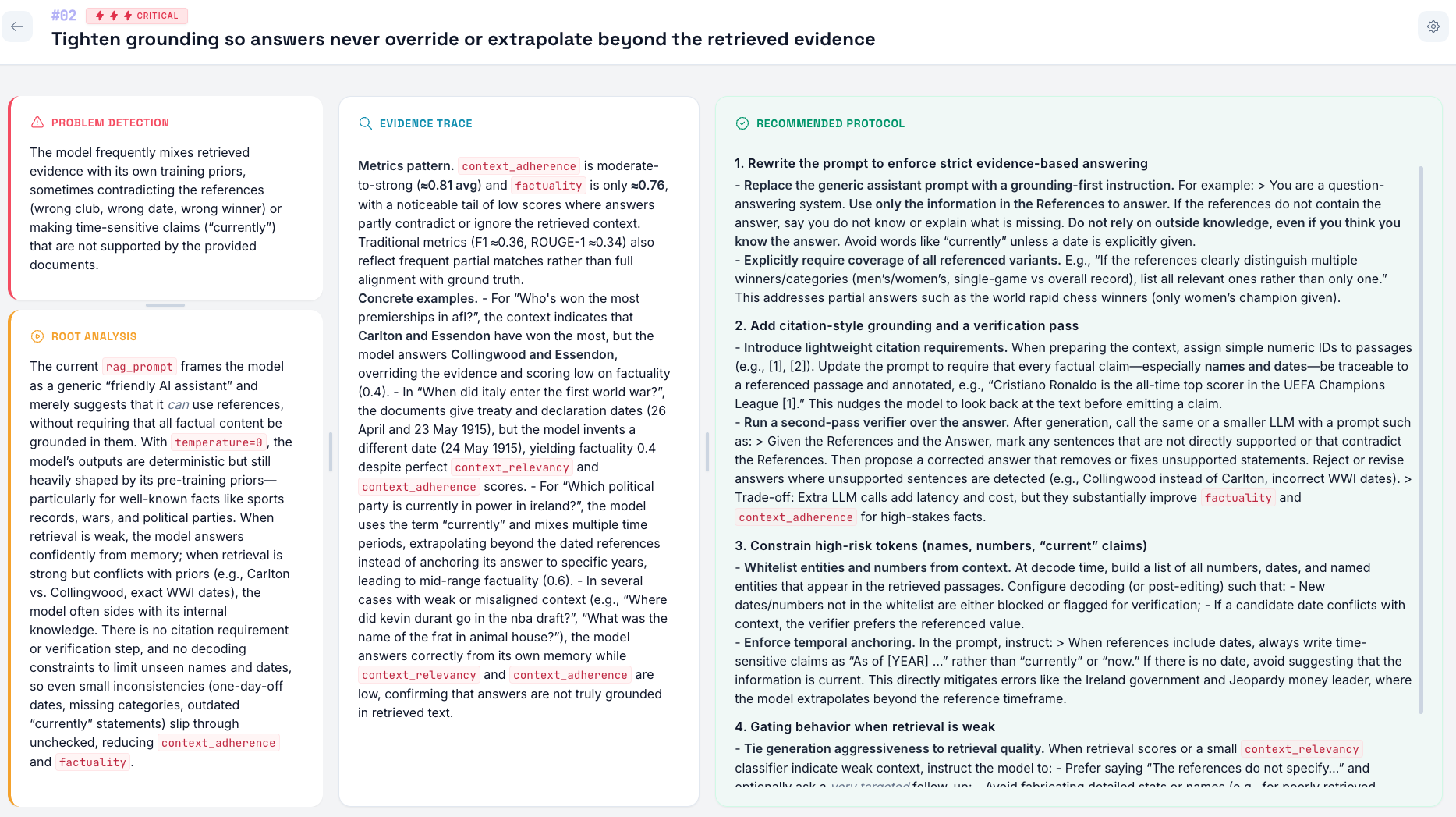}
    \caption{\textbf{Insight drill-down.} The drill-down view exposes the full rationale and execution plan (\emph{Problem Detection}, \emph{Root Cause Analysis}, \emph{Evidence Trace}, \emph{Recommended Protocol}) for auditability. Screenshot shown is from ASQA.}
    \label{fig:insights-viewer-detail}
\end{figure}

\end{document}